\documentclass{jfm}
\usepackage{amssymb,amsmath,dsfont}     
\usepackage{soul}
\usepackage{multirow}
\usepackage{comment}
\usepackage{amsfonts} 
\usepackage{color,graphicx,float}
\usepackage{subcaption}
\usepackage{enumitem}
\usepackage{fancyhdr}
\usepackage{hyperref}
\usepackage{bm}
\usepackage{pdfpages}

\def\lambAB{\Lambda_{\text{I} \rightarrow \text{II}}}

\def\lambBA{\Lambda_{\text{II} \rightarrow \text{I}}}

\def\hmax{h_{\text{max}}}


\title{Dynamics of thin film flows on a vertical fibre with vapor absorption}

\author{Souradip Chattopadhyay\aff{1}\thanks{These authors contributed equally to this work.}, Zihao Yu\aff{1}\footnotemark[1], \\
Y. Sungtaek Ju\aff{2} \and Hangjie Ji\aff{1}\corresp{ \email{hangjie\_ji@ncsu.edu}}}

\affiliation{\aff{1}Department of Mathematics, North Carolina State University, Raleigh, NC 27695, USA
\aff{2}Department of Mechanical and Aerospace Engineering, University of California Los Angeles, CA 90095, USA}

\begin{document}
\maketitle

\begin{abstract}
Water vapor capture through free surface flows plays a crucial role in various industrial applications, such as liquid desiccant air conditioning systems, water harvesting, and dewatering. This paper studies the dynamics of a silicone liquid sorbent (also known as water-absorbing silicone oil) flowing down a vertical cylindrical fibre while absorbing water vapor. We propose a one-sided thin-film-type model for these dynamics, where the governing equations form a coupled system of nonlinear fourth-order partial differential equations for the liquid film thickness and oil concentration. The model incorporates gravity, surface tension, Marangoni effects induced by concentration gradients, and non-mass-conserving effects due to absorption flux. Interfacial instabilities, driven by the competition between mass-conserving and non-mass-conserving effects, are investigated via stability analysis. We numerically show that water absorption can lead to the formation of irregular wavy patterns and trigger droplet coalescence downstream. Systematic simulations further identify parameter ranges for the Marangoni number and absorption parameter that lead to the onset of droplet coalescence dynamics and regime transitions.
\end{abstract}


\section{Introduction}
\label{sec:1}

Thin liquid films flowing down vertical cylinders have been extensively studied due to their wide range of technological applications, including fibre coating \citep{quere1999fluid}, textiles \citep{minor1959migration,patnaik2006wetting}, inkjet printing \citep{lohse2022fundamental}, and various other fields \citep{chinju2000string,binda2013integration}. Fibre coating, in particular, plays a critical role in industrial processes, as coating materials are used to protect and lubricate surfaces \citep{quere1997liquid}. Beyond their practical relevance, these films display complex and fascinating interfacial dynamics, including the formation of droplets and traveling wave patterns \citep{gau1999liquid,kalliadasis1994drop,quere1990thin}. Unlike thin films on planar substrates, thin films on cylindrical substrates are inherently unstable due to the additional azimuthal curvature of the free surface. The Rayleigh-Plateau (RP) instability, driven by gravity and surface tension, leads to a wide variety of spatial and temporal dynamics in these systems. Furthermore, coherent structures, such as droplet-like pulses and bound states, can be observed in viscous films flowing down vertical fibres \citep{duprat2009liquid}. 

\par 
The experimental study by \cite{quere1990thin} identified a critical thickness for fibre coating flows below which solitary wave solutions occur. It was observed that this critical thickness scales with the cube of the fibre radius for the formation of a drop. 
Following this experimental work, various long-wave models have been developed to analyze the evolution of the film interface into an undulating surface, leading to the formation of traveling waves and droplets along vertical cylinders.
These include thin film models developed by \cite{trifonov1992steady} and \cite{frenkel1992nonlinear}, and further studied by \cite{kalliadasis1994drop} and \cite{chang1999mechanism}; thick film models \citep{kliakhandler2001viscous}; asymptotic models \citep{craster2006viscous,ji2019dynamics}; integral boundary layer (IBL) models \citep{sisoev2006film}; and weighted-residual integral-boundary-layer (WRIBL) models \citep{ruyer2008modelling,ruyer2012wavy}. 
Thin liquid film models typically assume that the film thickness is much smaller than the cylinder radius, while asymptotic models treat the film radius as negligible compared to its characteristic length. Both approaches simplify the Navier-Stokes equations to a single evolution equation for the film thickness, making them well-suited for low Reynolds number flows. In contrast, the IBL and WRIBL models derive a system of evolution equations for both the film thickness and volumetric flow rates, making them applicable to thin-film flows with moderate Reynolds numbers. In addition, \cite{novbari2009energy} used the energy integral method to study the dynamics of an axisymmetric liquid film on a vertical cylinder at moderate Reynolds numbers. These models effectively capture a variety of dynamic phenomena in liquid films on vertical cylinders, including the formation of coherent structures and droplet coalescence \citep{duprat2007absolute,ji2021thermally}.

\par \cite{kliakhandler2001viscous} classified three flow regimes for axisymmetric sliding droplets on cylindrical fibres: convective, RP, and isolated. In the convective regime [regime (a)], faster-moving droplets collide with slower ones. In the RP regime [regime (b)], droplets travel with nearly constant inter-droplet spacing and speed, and coalescence does not occur. In the isolated droplet regime [regime (c)], the droplets are widely spaced, with smaller wavy patterns in between. Similar flow regimes have also been experimentally studied in asymmetric instabilities of thin-film flows along a vertical fibre by \cite{gabbard2021asymmetric}. 
The dependence of flow regimes and their transitions on system parameters has been identified using both lubrication-based and weighted-residual models \citep{craster2006viscous,ruyer2008modelling,ji2019dynamics}.

\par In addition to the studies mentioned earlier, other research has explored effects such as wall slip \citep{haefner2015influence,ding2011stability,halpern2017slip}, rotation \citep{rietz2017dynamics,liu2020instabilities,mukhopadhyay2020stability,chattopadhyay2025thermocapillary}, fibre geometry \citep{xie2021investigation,eghbali2022whirling,chao2024stability},
and the extension of dynamics to non-Newtonian liquids \citep{camassa2024long}.  
A significant portion of the literature has also explored the dynamics and stability of falling liquid films on vertical cylindrical fibres under thermocapillary effects, with a few notable studies by \cite{liu2017thermocapillary}, \cite{davalos2019sideband}, \cite{liu2014thermocapillary},  and \cite{ding2017three}. On the other hand, studies by  \cite{chattopadhyay2024modeling1}, \cite{chattopadhyay2024thermocapillary}, and  \cite{chao2020reactive} have explored the dynamics of reactive liquid films along a vertical fibre.
The studies by \cite{wray2013electrified,wray2013electrostatically} considered fibre coating dynamics in the presence of an electric field. 
Moreover, recent works by \cite{marzuola2020nonnegative},  \cite{ji2022travelling}, and
\cite{taranets2024weak} studied the well-posedness of PDE models for liquid films flowing down a cylinder. 
\cite{kim2024positivity} and \cite{biswal2024optimal} also studied positivity-preserving numerical schemes and optimal boundary control of fibre coating systems.

\par From a practical perspective, capturing water vapor has numerous engineering applications.  Liquid desiccant cooling, for example, is an energy-efficient and environmentally friendly option for air conditioning, especially under hot and humid climate conditions \citep{gurubalan2019comprehensive,chen2020recent,gurubalan2021comprehensive,fahlovi2023review}. The process of capturing water vapor is also critical in several freshwater generation techniques, such as harvesting ambient moisture, recovering vapor from cooling towers in thermoelectric power plants, and utilizing humidification-dehumidification (HDH) systems for desalination \citep{giwa2016recent,sadeghpour2019water}. Traveling liquid droplets generated through instability in thin film flow down vertical fibres and act as very effective radial mass sinks.  The enhanced mass transfer effectiveness, in turn, helps make more compact, economical, and energy-efficient dehumidifiers \citep{sadeghpour2019water}.

\par Most previous fibre coating models assume mass conservation, and the dynamics of non-conservative systems remain poorly understood.  In this study, we investigate a fibre coating model in which mass is not conserved due to water vapor absorption from the surrounding environment. We recognize a significant gap in the literature regarding the theoretical understanding of flow dynamics in thin liquid films along a vertical fibre during vapor absorption.  Past studies have explored vapor condensation and fluid evaporation in thin films, but only on (near) planar or inclined substrates. 
For example, \cite{burelbach1988nonlinear} developed a one-sided model for evaporating or condensing liquid films spreading on a uniformly heated or cooled horizontal plane, where the dynamics of the vapor phase were decoupled from those of the liquid phase. \cite{oron1999dewetting,oron2001dynamics} analyzed the dynamics of condensing thin films influenced by intermolecular forces on both horizontal and inclined substrates. 
\cite{bourges1995influence} provided experimental insights into the dynamics of volatile droplets, and \cite{oliver1968mass} conducted experiments to understand the influence of carbon dioxide and oxygen absorption by water films.  \cite{ajaev2005spreading},  \cite{shklyaev2007stability}, \cite{ji2018instability}, among others, investigated the instability of volatile liquid films and droplets.

\par For thin film flows down vertical fibres, vapor absorption is expected to strongly influence the flow regimes \citep{kliakhandler2001viscous} and transitions among them. \cite{sadeghpour2017effects} and  \cite{ji2020modelling} demonstrated that the flow regime is a sensitive function of the flow inlet conditions as well as flow rates. These, in turn, govern the droplet's size, speed, and spacing.  Another key aspect is identifying the factors that can trigger droplet collisions. For example,  \cite{ji2021thermally} found that applying a streamwise thermal gradient can trigger droplet coalescence.  A comprehensive understanding of droplet behavior is important as it governs the performance of heat and mass transfer and particle capture \citep{sadeghpour2021experimental}.

\par The paper is organized as follows: In section \ref{sec:2}, we propose a new lubrication model for a thin liquid flowing down a vertical fibre, where the liquid absorbs water vapor from the surrounding environment. Initially, the liquid is assumed to be purely a silicone liquid sorbent (also known as water absorbing silicone oil), which begins to flow along the fibre. A detailed characterization of this liquid is provided during parameter estimation in section \ref{sec:3b}.  As time progresses, the silicone liquid sorbent moves downward under the influence of gravity while simultaneously absorbing water vapor. After a certain period, the liquid mixture becomes saturated, and no further absorption occurs. Section \ref{sec:3} derives the model equations for the film thickness and silicone liquid sorbent concentration. In section \ref{sec:4}, we perform a linear stability analysis of the model, focusing on time-dependent base states for both film thickness and oil concentration. In section \ref{sec:5}, we present numerical simulations of the model to investigate various aspects of the dynamics influenced by water vapor absorption and Marangoni effects, including droplet coalescence and regime transitions, with concluding remarks provided in section \ref{sec:6}.

\section{Mathematical model}\label{sec:2}
\begin{figure}
\centering
\includegraphics[scale=0.8]{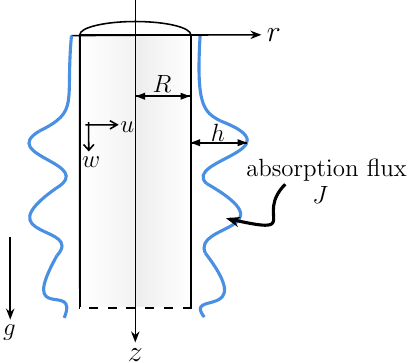}
\caption{Schematic of a thin liquid film down a vertical fibre with water vapor absorption flux $J$.}
\label{fig:schematic}
\end{figure}

We consider a two-dimensional axisymmetric flow of a thin film of silicone liquid sorbent along the outer surface of a vertical fibre of radius $R$ (see Figure \ref{fig:schematic}). Due to vapor absorption, the liquid is a mixture of silicone liquid sorbent and water. Here, we use cylindrical coordinates $(r,z)$ to describe the geometry, where the radial $r$ direction is perpendicular to the fibre axis, and the streamwise $z$ axis is directed downward along the fibre axis. We use $c(r,z,t)$ and $\eta(r,z,t)=1-c(r,z,t)$ to represent the concentration of silicone liquid sorbent and water in the liquid mixture, respectively, where $t$ denotes time. The liquid-air interface is denoted by $r=R+h$, where $h=h(z,t)$ represents the film thickness at time $t$. Below, we lay out dimensional governing equations and boundary conditions for the coupled dynamics of the film flow and the mixture concentration.

\subsection{Model formulation}
\par 
The dynamics of the liquid flow is governed by the continuity and Navier–Stokes equations
\begin{subequations}\label{dim_model}
\begin{equation}\label{eq2}
r^{-1}\left(ru\right)_r+w_z=0,
\end{equation}
\begin{equation}\label{eq3}
\rho\left(u_t+uu_r+wu_z\right)=-p_r+\mu\left[r^{-1}\left(ru_r\right)_r+u_{zz}-r^{-2}u\right],
\end{equation}
\begin{equation}\label{eq4}
\rho\left(w_t+uw_r+ww_z\right)=-p_z+\mu\left[r^{-1}\left(rw_r\right)_r+w_{zz}\right]+\rho g,
\end{equation}
where $\rho$ is the density, $\mu$ is the dynamic viscosity, $g$ is the gravitational acceleration, $p$ represents the pressure, and $u$ and $w$ represent the velocity components in the radial and axial directions, respectively.

\par The transport and diffusion of silicone liquid sorbent in the liquid mixture is described by the advection-diffusion equation
\begin{equation}\label{eq5}
c_t+\left(uc\right)_r+\left(wc\right)_z=d_m\left[r^{-1}\left(rc_r\right)_r+c_{zz}\right],
\end{equation}
where the constant $d_m$ is the molecular diffusivity.

\par At the fibre wall $r=R$, we impose  the usual no-slip and no-penetration conditions
\begin{equation}\label{eq6}
u=0, \quad w=0.
\end{equation}
At the free surface $r=R+h(z,t)$, that is, at the liquid-air interface, the tangential and normal stress balances lead to the boundary conditions
\begin{equation}\label{eq7}
\mu\left[\left(u_z+w_r\right)\left(1-h_z^2\right)+2\left(u_r-w_z\right)h_z\right]=\left(\sigma_z+h_z\sigma_r\right)\left(1+h_z^2\right)^{1/2},
\end{equation}
\begin{multline}
\label{eq8}
p_a-p+2\mu\left[u_r-\left(u_z+w_r\right)h_z+w_zh_z^2\right]\left(1+h_z^2\right)^{-1}\\
=\sigma\left[h_{zz}\left(1+h_z^2\right)^{-1}-r^{-1}\right]\left(1+h_z^2\right)^{-1/2},
\end{multline}
where $p_a$ is the pressure in the atmosphere.

\par The kinematic boundary condition at the free interface $r=R+h(z,t)$ is
\begin{equation}\label{eq9}
u-h_t-wh_z=-J,
\end{equation}
where $J$ represents the net flux of water vapor absorbed at the liquid-air interface. The negative sign indicates that water vapor enters the liquid. The specific form of $J$ used in this study is presented in section \ref{sec:2b}.

\par At the liquid-air interface $r=R+h(z,t)$, 
the normal component of the silicone liquid sorbent's diffusive flux balances the water vapor absorption rate at the interface. This leads to the boundary condition at $r=R+h(z,t)$,
\begin{equation}\label{eq5n}
d_m\left(c_r - c_z h_z\right)\left(1+h_z^2\right)^{-1/2} = -Jc.
\end{equation}
Moreover, we impose no flux condition at the fibre wall $\left(r = R\right)$ as 
\begin{equation}\label{eq5n1}
c_r=0.
\end{equation}
\end{subequations}
\par We assume that the surface tension $\sigma$ linearly varies with the concentration of silicone liquid sorbent $c$ as:
\begin{equation}\label{eq1}
\sigma(c)=\sigma_{\text{water}}-c\left(\sigma_{\text{water}}-\sigma_{\text{oil}}\right),
\end{equation}
where the oil concentration $0 < c < 1$,
$\sigma_{\text{oil}}$ represents the surface tension of pure silicone liquid sorbent, and $\sigma_{\text{water}}$ is the surface tension of pure water. Since $\sigma_{\text{water}}>\sigma_{\text{oil}}$, increasing the concentration of silicone liquid sorbent reduces the surface tension of the liquid mixture. 
Equation \eqref{eq1} also ensures that as $c\to1$, the surface tension $\sigma$ approaches that of pure silicone liquid sorbent, $\sigma\to \sigma_{\text{oil}}$. On the other hand, as $c\to0$, we have $\sigma\to\sigma_{\text{water}}$. The other liquid properties, such as density $\rho$ and dynamic viscosity $\mu$, are assumed constant. 

\subsection{Absorption flux}\label{sec:2b}
In the present study, we consider silicone liquid sorbent in contact with water vapor in the air. The driving force for water absorption is the pressure difference, $p_v - p_{\text{ve}}$, where $p_v$ is the vapor pressure in the air above the free interface, and $p_{\text{ve}}$ is the saturated water vapor pressure within the liquid mixture. 
To model the absorption process, we assume the absorption flux $J$ is proportional to the pressure difference \citep{ji2023mathematical,karapetsas2016evaporation}:
\begin{equation}\label{eq_abs_flux00}
J\propto p_v - p_{\text{ve}}.
\end{equation}
In practice, the vapor pressure $p_v$ can be controlled by adjusting the humidity of the surrounding environment.
The interplay between $p_v$ and $p_{\text{ve}}$ determines the direction of mass transfer: water vapor is absorbed when $p_v > p_{\text{ve}}$, the system remains at equilibrium when $p_v = p_{\text{ve}}$, and desorption into the gas phase occurs when $p_v < p_{\text{ve}}$. In the present study, we focus on the case of water vapor absorption.

\par According to Henry's law \citep{henry1832experiments}, the concentration of gas (water vapor in this study) dissolved in a liquid is proportional to the partial pressure of the gas (water vapor). Denoting $\eta$ as the dissolved water concentration and $\eta_s$ as the dissolved water concentration at saturation in the liquid mixture, Henry's law yields: $\eta=\mathbb H p_{\text{ve}}$ and $\eta_s=\mathbb H p_v$, where $\mathbb H >0$ is Henry's constant. 
Consequently, we arrive at
\begin{equation}\label{eq_abs_flux001}
\mathbb H \left(p_v-p_{\text{ve}}\right)=\eta_s-\eta,\quad \mbox{where}~ \eta_s>\eta.
\end{equation}
The absorption process in the silicone liquid sorbent-water mixture continues as long as $p_v>p_{\text{ve}}$, that is, until $\eta<\eta_s$. The absorption stops once the dissolved water concentration in the liquid mixture reaches $\eta_s$. Using (\ref{eq_abs_flux001}), we express the absorption flux $J$ in (\ref{eq_abs_flux00}) as 
\begin{equation}\label{eq_abs_flux0a}
J=J_0\left(\eta_s-\eta\right),
\end{equation}
where $J_0$ is the mass transfer coefficient.  

\par Using the relationship between the oil concentration $c$ and the water concentration $\eta$ $(c = 1 - \eta)$, the flux $J$ in (\ref{eq_abs_flux0a}) can be written in terms of the oil concentration $c$ as:
\begin{equation}\label{eq_abs_flux}
J=J_0\left(c - c_s\right), \quad 0< c_s\leq c\leq 1,
\end{equation}
where $c_s = \left(1 - \eta_s\right)$ is the saturated oil concentration in the liquid mixture. This form of $J$ is used in the final model equation \eqref{eq:main}.

\subsection{Scalings and nondimensionalization}
We use the following scaling to derive the dimensionless governing equations and boundary conditions
\begin{align}
\label{eq11}
&z=\mathcal Lz^*,~(h,r,R)=\mathcal H(h^*,r^*,R^*),~ t=\left(\mathcal L/\mathcal V\right)t^*,~ (u,w)=\mathcal V\left(\epsilon u^*,w^*\right),\nonumber
\\
&p=p_a+\mathcal P p^*,~J=J_0J^*,~\sigma=\sigma_0\sigma^*,
\end{align}
where the quantities marked with an asterisk denote dimensionless quantities. Here, $\mathcal H$ is the mean thickness of the film, which we consider as the length scale along the radial direction, $\mathcal V$ is the velocity scale, and $\mathcal P$ is the pressure scale.
The surface tension $\sigma$ is scaled by $\sigma_0$, where $\sigma_0$ represents the surface tension of the silicone liquid sorbent at the reference temperature, i.e.,  $\sigma_0=\sigma_{\text{oil}}$. 
We choose the length scale in the axial direction $z$ as $\mathcal L=\mathcal H/\epsilon$. Following \cite{ji2019dynamics,ji2021thermally}, in this model, we set the aspect ratio $\epsilon$ by balancing the surface tension and the gravity $g$, which leads to $\epsilon=\left(\rho g \mathcal H^2/\sigma_0\right)^{1/3}$. We set the pressure scale $\mathcal P=\rho g\mathcal L$ and the velocity scale $\mathcal V=g\mathcal H^2/\nu$, where $\nu=\mu/\rho$ is the kinematic viscosity. 

\par Using (\ref{eq11}) in (\ref{dim_model}), we obtain the following system of equations in dimensionless form. For simplicity, we have dropped the asterisk in the dimensionless variables.
\begin{subequations}\label{dimmless_model}
\begin{equation}\label{eq12}
r^{-1}\left(ru\right)_r+w_z=0,
\end{equation}
\begin{equation}\label{eq13}
\epsilon^4Re\left(u_t+uu_r+wu_z\right)=-p_r+\epsilon^2
\left[r^{-1}\left(ru_r\right)_r+\epsilon^2u_{zz}-r^{-2}u\right],
\end{equation}
\begin{equation}\label{eq14}
\epsilon^2Re\left(w_t+uw_r+ww_z\right)=1-p_z+
r^{-1}\left(rw_r\right)_r+\epsilon^2w_{zz},
\end{equation}
\begin{equation}\label{eq15}
c_t+\left(uc\right)_r+\left(wc\right)_z=\delta\left[\epsilon^{-2}r^{-1}\left(rc_r\right)_r+c_{zz}\right],
\end{equation}
\begin{equation}\label{eq16}
u=0, \quad w=0, \quad c_r=0 \quad \text{at}~r=R, 
\end{equation}
\begin{equation}\label{eq17}
\left(\epsilon^2u_z+w_r\right)\mathcal E+2\epsilon^2\left(u_r-w_z\right)h_z=M\left(\sigma_z+h_z\sigma_r\right)\mathcal F \quad \text{at}~r=R+h(z,t),
\end{equation}
\begin{multline}\label{eq18}
-p+2\epsilon^2\left[u_r-\left(\epsilon^2u_z+w_r\right)h_z+\epsilon^2w_zh_z^2\right]\mathcal F^{-2}\\
=\epsilon^3We\left[h_{zz}\left(1+\epsilon^2h_z^2\right)^{-1}-\epsilon^{-2}r^{-1}\right]\sigma\mathcal F^{-1} \quad \text{at}~r=R+h(z,t),
\end{multline}
\begin{equation}\label{eq19}
u-h_t-wh_z=-\Lambda J \quad \text{at}~r=R+h(z,t),
\end{equation}
\begin{equation}\label{eq15n11}
c_r - \epsilon^2 \left(c_z h_z -\Lambda\delta^{-1}Jc\mathcal F\right)=0 \quad \text{at}~r=R+h(z,t),
\end{equation}
\end{subequations}
where $Re=\mathcal{VL}/\nu$ is the Reynolds number,  $M=\sigma_0/\left(\rho g\mathcal {HL}\right)$ is the Marangoni number,   $\Lambda=\mathcal LJ_0/\left(\mathcal {HV}\right)$ is the absorption rate, $Pe=\mathcal{VL}/d_m$ is the Peclet number, $\delta=Pe^{-1}$, $\mathcal E=1-\epsilon^2h_z^2$, and $\mathcal F=\left(1+\epsilon^2h_z^2\right)^{1/2}$. 
The Weber number $We$ is defined as the square of the ratio of the capillary length $l_c$ to the radial length scale $\mathcal H$, where $l_c = \sqrt{\sigma_0/(\rho g)}$. Following \cite{ji2019dynamics}, we adopt the scale ratio $\epsilon = We^{-1/3}$ and assume $\delta=O(1)$ in the current model. Consequently, the term $\epsilon^3We = 1$ in (\ref{eq18}).

\par In (\ref{dimmless_model}), the scaled surface tension $\sigma$ is obtained by substituting (\ref{eq11}) into (\ref{eq1}) as
\begin{equation}\label{eq1a}
\sigma(c)=1+K_\sigma\eta = 1 + K_{\sigma}(1-c), \quad K_{\sigma} > 0,
\end{equation}
where $\eta=1-c$ is the concentration of water in the liquid mixture and $K_\sigma=(\sigma_{\text{water}}-\sigma_{\text{oil}})/\sigma_{\text{oil}}$ is a positive constant since $\sigma_{\text{water}}>\sigma_{\text{oil}}$.

\par Under the lubrication approximation, we neglect the inertial contributions by assuming that $Re= O(1)$ and $\epsilon\ll1$ \citep{craster2006viscous}. Dropping the terms of $O\left(\epsilon^2\right)$ and higher from (\ref{eq13})-(\ref{eq14}) and (\ref{eq16})-(\ref{eq19}) yields the following leading order system of equations:
\begin{subequations}\label{lead_eqs}
\begin{equation}\label{eq21}
p_r=0,\quad
 r^{-1}\left(rw_r\right)_r=p_z-1,
\end{equation}
\begin{equation}\label{eq22}
u=0, \quad w=0 \quad \text{at}~r=R,
\end{equation}
\begin{align}\label{eq23}
&w_r=M\left(\sigma_z+h_z\sigma_r\right), \quad  
p=\sigma(c)\left(\epsilon^{-2}r^{-1}-h_{zz}\right),\nonumber
\\
\quad 
&u=h_t+wh_z-\Lambda J \quad \text{at}~r=R+h(z,t).
\end{align}
\end{subequations}

\section{Cross-sectional averaging and asymptotic model}
\label{sec:3}
Following the approach by \cite{jensen1993spreading} on modeling the spreading of heat and soluble surfactant in thin liquid films, 
we express the concentration $c$ by decomposing it into an averaged component independent of $r$ and a small fluctuation,
\begin{equation}\label{eqn_1}
c=c_0(z,t)+\epsilon^2\delta^{-1}c_1(r,z,t),
\end{equation}
where the cross-sectional average of the fluctuation $c_1(r,z,t)$ is zero,
\begin{equation}\label{eqn_2}
\widehat c_1(r,z,t)\equiv\frac{1}{R+h}\int_R^{R+h}c_1(r,z,t)r\text{d}r=0.
\end{equation}

\par We define the flow rate $q$ as $q=\int_R^Srw\text{d}r$ and $S(z,t)=R+h(z,t)$. Inserting (\ref{eqn_1}) into (\ref{eq15}) and averaging the obtained equation with respect to $r$ using  (\ref{eqn_2}) yields the following form after omitting terms of $O\left(\epsilon^2\right)$ and higher:
\begin{multline}\label{eqn_4_n11}
\left[c_0\left(S^2-R^2\right)\right]_t+2\left(qc_0\right)_z\\
=2\left(rc_{1,r}\right)_R^S+\delta\left[c_{0,z}\left(S^2-R^2\right)\right]_z-2S\delta c_{0,z}h_z+2c_0\left(Sh_t+q_z\right). 
\end{multline}
 
\par Assuming the concentration only varies in the axial direction, we obtain the expression for the streamwise velocity $w$ by solving the system of equations (\ref{lead_eqs}) as
\begin{equation}\label{eq26}
w=\left(1-p_z\right)\left[\frac{S^2}{2}\ln\left(\frac{r}{R}\right)+\frac{\left(R^2-r^2\right)}{4}\right]-K_\sigma MS\ln\left(\frac{r}{R}\right)c_z,
\end{equation}
and consequently, obtain the flow rate $q$ as 
\begin{multline}\label{eq26a}
q=\left(1-p_z\right)\left[\frac{S^4}{4}\ln\left(\frac{S}{R}\right)+\frac{\left(3S^2-R^2\right)\left(R^2-S^2\right)}{16}\right]\\
-K_\sigma MS\left[\frac{S^2}{2}\ln\left(\frac{S}{R}\right)+\frac{R^2-S^2}{4}\right]c_z.
\end{multline}

\par Substituting \eqref{eq26a} into the mass conservative form of the kinematic boundary condition (\ref{eq19}) yields
\begin{subequations}\label{eq13aa}
\begin{equation}
\left(1+\alpha h\right)h_t+q_z=\left(1+\alpha h\right)\Lambda J,
\end{equation}
where 
\begin{equation}\label{eq13aa_flux}
q=\frac{h^3}{3}\left(1-p_z\right)\phi(\alpha h)-\frac{M}{2}h^2\psi(\alpha h)K_\sigma c_z
\end{equation}
and $\alpha=1/R$. The parameter $\alpha$ also represents the aspect ratio between the characteristic dimensional length scale in the radial direction to the dimensional fibre radius. 

\par The shape factors $\phi(Y)$ and $\psi(Y)$ in (\ref{eq13aa_flux}) take the following forms:
\begin{equation}
\phi(Y)=\frac{3\left[(1+Y)^4\left(4\ln(1+Y)-3\right)+4(1+Y)^2-1\right]}{16Y^3},
\end{equation}
\begin{equation}
\psi(Y)=\frac{(1+Y)}{Y^2}\left[(1+Y)^2\ln(1+Y)-Y\left(1+\frac{Y}{2}\right)\right].
\end{equation}
In the limit of $Y\rightarrow 0$,
\begin{equation}\label{sskm}
\phi(Y)=1+Y+\frac{3Y^2}{20}+O\left(Y^3\right),\quad \psi(Y)=1+\frac{4Y}{3}+\frac{Y^2}{4}+O\left(Y^3\right).
\end{equation}

\par The pressure solution in the flow rate $q$ is obtained by solving the system of equations (\ref{lead_eqs}), yielding 
\begin{equation}\label{eq25}
p=\frac{1}{\varsigma}\left(\frac{\alpha}{(1+\alpha h)}-\varsigma h_{zz}\right),
\end{equation}
\end{subequations}
where $\varsigma=\epsilon^2$.  
Here, we approximate $\sigma(c)\approx 1$ in the expression for the dynamic pressure in \eqref{eq23}, since for the silicone liquid sorbent considered in this study, the saturated water concentration is close to $14\%$ \citep{ahn2017method} and does not significantly affect the bulk surface tension of the liquid mixture.
We retain the streamwise curvature term $\varsigma h_{zz}$ in the leading-order equation because the curvature of the liquid/vapor interface can be large in certain regions, and omitting the stabilizing streamwise curvature term may lead to ill-posedness in the lubrication model. Previous studies on similar geometries by \cite{craster2006viscous}, \cite{liu2018effect}, and many others have also considered this term in their analysis.

\par Substituting (\ref{eqn_1}) into (\ref{eq15n11}) yields
\begin{equation}\label{eqq_a1}
\left(c_{1,r}-\delta c_{0,z}h_z\right)_{r=S}=-\Lambda Jc_0,\quad \left.c_{1,r}\right|_{r=R}=0.
\end{equation}
Using (\ref{eq13aa}) and (\ref{eqq_a1}) in (\ref{eqn_4_n11}) leads to the governing equation for the cross-sectional average concentration $c_0$,
\begin{multline}\label{eqn_4_nn112}
\left[c_0\left(h+\frac{\alpha}{2}h^2\right)\right]_t+\left[\left(\frac{h^3}{3}\left(1-p_z\right)\phi(\alpha h)-Mah^2\psi(\alpha h)c_z\right) c_0\right]_z\\
=\delta\left[c_{0,z}\left(h+\frac{\alpha}{2}h^2\right)\right]_z, 
\end{multline}
where $Ma=K_\sigma M/2$ is the modified Marangoni number. Equation \eqref{eqn_4_nn112} characterizes the contributions from gravity, the bulk surface tension, the Marangoni effects induced by the concentration gradient, and the diffusion to the dynamics of the oil concentration.

\subsection{Nonlinear evolution equations of film thickness and oil  concentration}\label{sec:3a}
We replace $c_0$ by $c$ in \eqref{eqn_4_nn112} for notational simplicity and introduce a change of time scale $t \to t/\phi(\alpha)$.  
Finally, we obtain the coupled dimensionless PDE system for the liquid thickness $h(z,t)$ and the concentration of the silicone liquid sorbent $c(z,t)$ as 
\begin{subequations}
\label{eq:main}
\begin{equation}\label{ss1}
\left(h+\frac{\alpha}{2}h^2\right)_t+q_z=\Lambda\left(1+\alpha h\right)J,  
\end{equation}
\begin{equation}\label{ss2}
\left[c\left(h+\frac{\alpha}{2}h^2\right)\right]_t+\left(qc\right)_z=\delta\left[c_z\left(h+\frac{\alpha}{2}h^2\right)\right]_z,
\end{equation}
\begin{equation}\label{ss3}
q=\mathcal M(h)\left[1-\left(\frac{\alpha}{\varsigma\left(1+\alpha h\right)}-h_{zz}\right)_z\right]-Mah^2\frac{\psi(\alpha h)}{\phi(\alpha)}c_z,
\end{equation}
where the mobility function $\mathcal M(h)$ is given by
\begin{equation}
\mathcal M(h) =  \frac{h^3}{3}\frac{\phi(\alpha h)}{\phi(\alpha)},
\end{equation}
and the dimensionless form of the absorption flux in (\ref{ss1}) is
\begin{equation}\label{swq1}
J(c) = \left(c-c_s\right) \ge 0, \quad  c_s \le c \le 1,
\end{equation}
\end{subequations}
where $c_s$ is the saturated concentration of silicone liquid sorbent in the liquid mixture and satisfies $0 < c_s < 1$. The right-hand side of \eqref{ss1} represents the mass flux due to water vapor absorption. 
The first term in \eqref{ss3} accounts for gravity, the second term $\alpha/[\varsigma(1+\alpha h)] - h_{zz}$ describes the dual role of surface tension due to a destabilizing azimuthal curvature term and a stabilizing streamwise curvature term, and the last term in \eqref{ss3} describes Marangoni effects induced by the concentration gradient in $c$. 

\par The dynamics described by the coupled PDE system \eqref{eq:main} are strongly influenced by boundary conditions. For the downstream interfacial flow in the Raleigh-Plateau regime away from the nozzle inlet, it is convenient to use periodic boundary conditions \citep{ji2019dynamics}. To study the full dynamics, including the near-nozzle flow patterns, it is necessary to incorporate the nozzle geometry and the flow rate in the boundary conditions \citep{ji2020modelling}. In section \ref{sec:4},  we will analyze the stability of the system \eqref{eq:main} over a periodic domain $0 \le z \le L$, with a focus on the instability triggered by absorption. In section \ref{sec:5}, we will numerically study the system with boundary conditions that are compatible with spatially-dependent absorption influenced by near-nozzle dynamics.

\par It is useful to define the total mass of the liquid mixture and the total mass of silicone liquid sorbent as
\begin{equation}
    M_l(t) = \int_0^L \left(h+\frac{\alpha}{2}h^2\right) \text{d}z, \quad M_o(t) = \int_0^L c\left(h+\frac{\alpha}{2}h^2\right) \text{d}z.
\label{eq:mass}
\end{equation}
For dynamics over a periodic domain $0\le z \le L$, integrating \eqref{ss1} and \eqref{ss2} gives the rates of change of the total masses as
\begin{equation}
\frac{d M_l}{dt} = \Lambda\int_0^L (1+\alpha h) J ~\text{d}z \ge 0, \quad \frac{dM_o}{dt} = 0,
\end{equation}
indicating that the total mass of liquid increases if $\Lambda > 0$ and $c > c_s$ due to vapor absorption, while the total mass of silicone liquid sorbent remains conserved over time.

\par We note that in the absence of Marangoni effects and vapor absorption ($Ma=\Lambda = 0$), the equations for $h$ and $c$ in \eqref{eq:main} become decoupled, and the governing equation for the film thickness becomes
\begin{equation}\label{eqrse}
\left(h+\frac{\alpha}{2}h^2\right)_t + \left[\mathcal M(h)\left(1+\frac{\alpha^2}{\varsigma\left(1+\alpha h\right)^2}h_z+h_{zzz}\right)\right]_z = 0.
\end{equation}
This is consistent with the classical lubrication model for viscous thin liquid flowing down a vertical fibre, which has been widely studied \citep{craster2006viscous,ji2019dynamics}. The incorporation of the non-mass-conserving effects due to the absorption flux and Marangoni effects into the system is expected to yield more interesting and complex dynamics. 
For $\alpha h \to 0$, equation (\ref{eqrse}) simplifies to the form derived by \cite{kalliadasis1994drop}, and in the absence of gravitational effects, it reduces to Hammond's equation \citep{hammond1983nonlinear}.

\subsection{Parameter estimation}\label{sec:3b}
In this section, we briefly discuss the system parameter ranges relevant to the current study. 
The present study considers silicone liquid sorbent, known as Dow XX-8810, which has been used in gas and vapor transport applications such as dehumidification \citep{ahn2017method}. This silicone liquid sorbent has density $\rho=1.031\times 10^3$kg m$^{-3}$, kinematic viscosity $\nu=3.492\times 10^{-5}$m$^2$s$^{-1}$, and surface tension $\sigma_0=22\times10^{-3}$N m$^{-1}$ at $20^\circ$C. 
At $20^\circ$C and $80\%$ relative humidity, the present silicone liquid sorbent can absorb up to $14.4\%$ of water vapor by weight, corresponding to a saturated silicone liquid sorbent concentration of $c_s\approx 0.86$.

\par We consider a typical fibre coating experiment where the fibre radius is $R = 3\times 10^{-4}$m and the dimensional volumetric flow rate $Q_m = 8\times 10^{-6}$kg s$^{-1}$. Following \cite{ruyer2008modelling} and \cite{ji2019dynamics}, we define the characteristic film thickness $\mathcal{H}$ by the Nusselt solution that satisfies $Q_m = \left[2\pi \rho R g \mathcal{H}^3\phi(\mathcal{H}/R)\right]/\left(3\nu\right)$. This leads to $\mathcal{H} = 2.8\times 10^{-4}$m, and the corresponding dimensionless system parameters $\alpha \approx 0.93$ and $\varsigma \approx 0.11$. For all numerical results in the rest of the paper, we set $\alpha = 0.93$, $\varsigma=0.11$, and the diffusion parameter $\delta = 0.01$.

\par The modified Marangoni number $Ma=K_\sigma M/2$, due to the concentration gradient, and the absorption parameter $\Lambda$ are two key system parameters that quantify the relative importance of Marangoni effects and non-mass-conserving effects. Since the surface tension of water at $20^\circ$C is $\sigma_{\text{water}} = 72.74\times10^{-3}$N m$^{-1}$, we have $K_\sigma=(\sigma_{\text{water}}-\sigma_{0})/\sigma_{0} \approx 2.3$. Correspondingly, the modified Marangoni number $Ma \approx 11$, and we explore the ranges $Ma\in[0,30]$ for theoretical analysis. The absorption parameter $\Lambda$ is more difficult to estimate due to the uncertainty in the $J_0$ scale in  (\ref{eq_abs_flux}), which represents the characteristic mass transfer velocity at which water molecules are absorbed into the silicone liquid sorbent. In this study, we consider both weak and moderate absorption rates by setting $\Lambda\in[0,5]$ and explore how water vapor absorption affects droplet dynamics.

\section{Linear stability analysis}\label{sec:4}
In this section, we perform the linear stability analysis of the liquid-air interface using the model \eqref{eq:main} to investigate the impact of water vapor absorption on the system. This section employs periodic boundary conditions along the axial direction. However, in section \ref{sec:5}, for the numerical simulation of the full model, we apply Dirichlet boundary conditions at the inlet and Neumann boundary conditions at the outlet, which are more physically appropriate for capturing the system's dynamics.

\par We expand the film thickness $h(z,t)$ and the oil concentration $c(z,t)$ around their base states $\overline h$ and $\overline c$,
\begin{equation}\label{eqs_1}
h(z,t)=\overline h(t)+\widetilde h(z,t),\quad c(z,t)=\overline c(t)+\widetilde c(z,t),
\end{equation}
where the tilde denotes the perturbed quantity. We will discuss two separate cases based on the properties of the base states. In subsection \ref{sec:4a}, we consider the case 
where the base state $\bar{c}$ remains at the saturated concentration, $\bar{c} \equiv c_s$, which corresponds to the scenario without vapor absorption. In subsection \ref{sec:4b}, we discuss the case of slow absorption with $\overline{c} > c_s$, using the frozen time approach \citep{shklyaev2007stability,burelbach1988nonlinear,ji2018instability} by assuming quasi-static base states $\bar{h}$ and $\bar{c}$.

\par While the saturated silicone liquid sorbent concentration $c_s$ was found to be approximately $0.86$ in section~\ref{sec:3b}, we choose a lower value of $c_s$ for the stability analysis. 
When $c_s$ is close to the initial concentration $c = 1$, the liquid film rapidly reaches saturation, and the influence of water absorption on the stability of the system is not prominent. To allow absorption to persist over a longer period and enhance instability driven by water absorption, we consider a smaller saturated concentration $c_s = 0.2$ instead. For the numerical studies of the PDE \eqref{eq:main} in section \ref{sec:5}, we revert to the estimated value $c_s = 0.86$.

\subsection{Stability with saturated water concentration}\label{sec:4a}
\par For the case when the base state $\overline c$ is identical to the saturated oil concentration $c_s$, $\overline c \equiv c_s$, from \eqref{swq1} we have the absorption flux $J(\bar{c}) = 0$. Also, from \eqref{swq1}, the total amount of liquid mixture is conserved in this case, $M_l(t) = M_l(0)$, indicating that the base state of the liquid film thickness $\bar{h}$ remains constant over time.

\par For simplicity, we set the base state $\overline h=1$ and express the perturbation as an infinitesimal Fourier mode disturbance
\begin{equation}\label{eqs_3}
\widetilde h(z,t)=\widehat h\exp\left[ikz+\Omega(t)\right],
\end{equation}
where $k$ is the wavenumber, $\Omega(t)$ is the growth rate of the perturbation starting from a small initial amplitude $\widehat h \ll 1$. Inserting $h=1+\widetilde h$ into (\ref{eq:main}) and linearizing around the base state $\bar{h}=1$ yields the dispersion relation for the growth rate $\Omega(t) = \Omega_r(t) + i\Omega_i(t)$, where 
the real and imaginary parts of the dispersion relation are obtained as
\begin{equation}\label{eqs_4}
\frac{d\Omega_r}{dt}=\frac{k^2}{3(1+\alpha)}\left(\frac{\alpha^2}{\varsigma(1+\alpha)^2}-k^2\right), \quad 
\frac{d\Omega_i}{dt}=-\frac{3\phi(\alpha)+\alpha\phi'(\alpha)}{3(1+\alpha)\phi(\alpha)}k,
\end{equation}
where prime denotes the derivative.

\par This is consistent with the linear stability results of the uniform Nusselt solution for liquid flowing down a vertical fibre previously studied by \cite{craster2006viscous}. The first equation in (\ref{eqs_4}) gives the linear growth rate, and the second describes the nondispersive linear wave speed, $c_L=\left[3\phi(\alpha)+\alpha\phi'(\alpha)\right]/\left[3(1+\alpha)\phi(\alpha)\right]$.  Equation (\ref{eqs_4}) also shows that when the silicone liquid sorbent is saturated with water vapor, the growth rate and wave speed resemble those of a simple gravity-driven liquid film along a vertical fibre. It also recovers the results of \cite{ji2019dynamics} when wall slipperiness and additional stabilization are neglected in their study. Additionally, (\ref{eqs_4}) predicts a critical wavenumber $k_c=\alpha/\left[\sqrt{\varsigma}(1+\alpha)\right]$ by setting the first equation of (\ref{eqs_4}) to zero, below which the flow is linearly unstable.  On the other hand,  for $k>k_c$, the flow is linearly stable. This critical wavenumber corresponds to the Rayleigh-Plateau instability. Moreover, we obtain the fastest growing mode $k_m=\alpha/\left[\sqrt{2\varsigma}(1+\alpha)\right]$.

\subsection{Stability with weak water vapor absorption}\label{sec:4b}
For the scenario with weak absorption ($\Lambda \ll 1$) and a base state for oil concentration $\bar{c} > c_s$, vapor absorption occurs, leading to a slowly increasing film thickness $(h)$ and decreasing in the oil concentration $(c)$ over time $t$. In this case, we express the time-dependent base states $\overline h$ and $\overline c$ in \eqref{eqs_1} as $\overline h=h_b(t)$ and $\overline c=c_b(t)$. To analyze the stability of the free surface under weak absorption, we assume that the base states are quasi-static and adopt the frozen time approach \citep{shklyaev2007stability,burelbach1988nonlinear,ji2018instability}, with the solutions to \eqref{eq:main} taking the form
\begin{equation}\label{eqs_5}
h(z,t)=h_b(t)+\chi\widehat{h}\exp\left[ikz+\lambda_h(t)\right], \quad c(z,t)=c_b(t)+\chi\widehat{c}\exp\left[ikz+\lambda_c(t)\right],
\end{equation}
where $\chi\ll1$, $\chi\widehat{h}$ and $\chi{\widehat{c}}$ are the initial amplitudes of the perturbation, $\widehat{h},\widehat{c}=O(1)$, $\lambda_h(t)$ and $\lambda_c(t)$ represent the growth of the perturbation satisfying $\lambda_h(0)=\lambda_c(0) = 0$, and $c_s< c_b(0) \le 1$.

\par Substituting (\ref{eqs_5}) into (\ref{eq:main}) and linearizing the system about the base states gives the $O(1)$ and $O(\chi)$ equations,
\begin{subequations}\label{eqs_6c}
\begin{equation}\label{eqs_6c_jj}
O(1): \qquad \frac{dh_b}{dt}=\Lambda\left(c_b-c_s\right), \quad c_b\left(h_b+\frac{\alpha}{2}h_b^2\right)=\left<M_o\right>,
\end{equation}

\begin{multline}\label{eqs_6b_n2a}
O(\chi):\quad \frac{d\lambda_h}{dt}+i\mathbb Ek+\\
\left[\dfrac{h_b^3\phi\left(\alpha h_b\right)}{3\phi(\alpha)\left(1+\alpha h_b\right)}\left(k^2- \frac{\alpha^2}{\varsigma\left(1+\alpha h_b\right)^2}\right)
+\frac{Mah_b^2\psi\left(\alpha h_b\right)}{\phi(\alpha)\left(1+\alpha h_b\right)}\frac{\widehat c}{\widehat h}\exp(\lambda_c - \lambda_h)\right]k^2
=\Lambda \Gamma,
\end{multline}
\begin{multline}\label{eqs_7a_n2}
\frac{d\lambda_c}{dt}+\frac{c_b\Lambda }{\left(h_b+\dfrac{\alpha}{2}h_b^2\right)}\left[1+\alpha h_b+\alpha\left(c_b-c_s\right)\frac{\widehat h}{\widehat c}\exp(\lambda_h-\lambda_c)\right]\\
+\frac{ik\mathcal M\left(h_b\right)}{\left(h_b+\dfrac{\alpha}{2}h_b^2\right)}+\delta k^2=0.
\end{multline}
\end{subequations}
\par In (\ref{eqs_6c}), $\left<M_o\right>$ is the average mass of silicone liquid sorbent per unit length along the domain, i.e., $\left<M_o\right>=M_o/L$, where the total mass of silicone liquid sorbent $M_o$ in the system is defined in \eqref{eq:mass} and 
\begin{equation}\label{gam}
\Gamma=\frac{\alpha\left(c_b-c_s\right)}{1+\alpha h_b}+\frac{\widehat c}{\widehat h}\exp(\lambda_c - \lambda_h),
\quad
\mathbb E=\frac{3h_b^2\phi\left(\alpha h_b\right)+\alpha h_b^3\phi'\left(\alpha h_b\right)}{3\phi(\alpha)\left(1+\alpha h_b\right)}.
\end{equation}

\par The $O(1)$ equations \eqref{eqs_6c_jj} govern the dynamics of the base states $h_b$ and $c_b$ over time. During the absorption process, the leading-order concentration profile $c_b$ gradually decreases and approaches the saturated concentration $c_s$ as $t\to \infty$, while the leading-order film thickness $h_b$ increases and approaches a terminal height $h_s$. Here, we have
$c_s(h_s+\alpha h_s^2/2) = \left<M_o\right>$ based on the conservation of oil mass. For all discussions and plots in section \ref{sec:4b}, we set the parameters
$c_s = 0.2$ and $\left<M_o\right> = 0.0512$.
\begin{figure}
\centering
\includegraphics[scale=0.25]{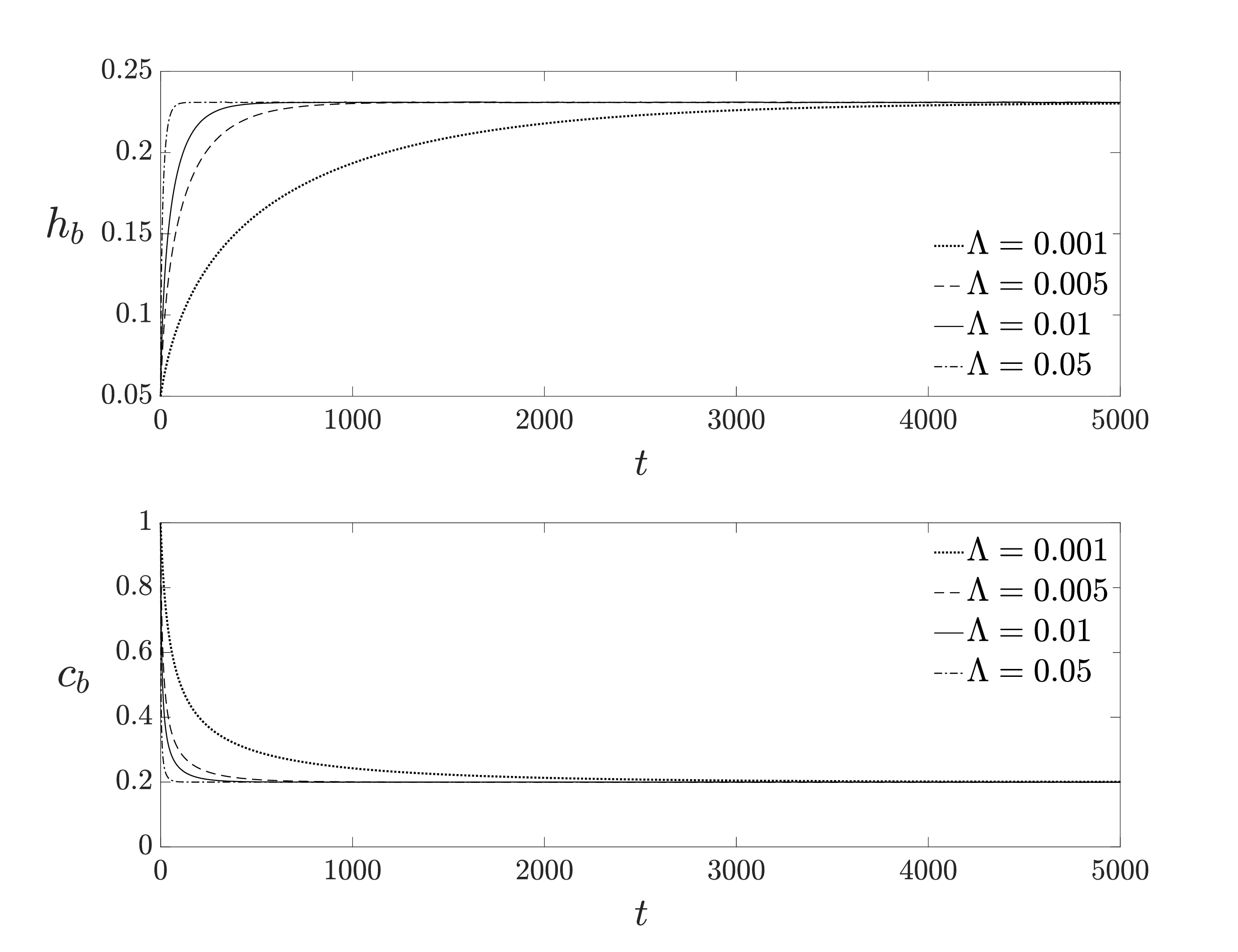}
\caption{Base state profiles from (\ref{eqs_6c_jj}) }
\label{fig:base}
\end{figure}

\par Figure \ref{fig:base} presents the typical evolution of the base state profiles for film thickness $\left(h_b\right)$ and oil concentration $\left(c_b\right)$ over time, obtained by numerically solving the $O(1)$ equations \eqref{eqs_6c_jj} with several values of the absorption parameter $\Lambda$.
With an initial concentration $c_b(0)=1$, the initial base state of the film thickness is determined from the second equation of (\ref{eqs_6c_jj}) as 
$h_b(0)=\left({-1+\sqrt{1+2\alpha \left<M_o\right>}}\right)/{\alpha}$. The plots in Figure \ref{fig:base} show that a larger absorption rate $\Lambda$ leads to a more rapid increase in the film thickness $h_b(t)$, while the concentration $c_b(t)$ decays correspondingly and approaches the saturated concentration $c_s=0.2$. 
For higher values of $c_s$, the total amount of absorbed water decreases, and the leading-order dynamics, governed by (\ref{eqs_6c_jj}), become less sensitive to $\Lambda$. 

\par From \eqref{eqs_7a_n2}, we observe that the real part of the growth rate of the perturbation in oil concentration is given by
\begin{equation}\label{eqs_7a_n21}
\frac{d\lambda_{c,r}}{dt}=-\left(\frac{c_b\Lambda }{\left(h_b+\dfrac{\alpha}{2}h_b^2\right)}\left[1+\alpha h_b+\alpha\left(c_b-c_s\right)\frac{\widehat h}{\widehat c}\exp(\lambda_{h,r} - \lambda_{c,r})\right]+\delta k^2\right) < 0.
\end{equation}
The growth rate is negative since $\Lambda$, $h_b$, $\alpha$, $\delta$ are all positive, and $0< c_s \le  c_b \le 1$. 
Therefore, with the initial exponent $\lambda_c(0)=0$, we have $\lambda_{c,r}(t) < 0$ for $t > 0$, that is, the magnitude of the perturbation in concentration is expected to decay in time. For nontrivial dynamics with pattern formation, we expect the effective growth of the perturbation in film thickness $\lambda_{h,r}$ to satisfy $\lambda_{h,r}(t) \ge 0 > \lambda_{c,r}(t)$, which indicates that the term with $Ma$ in \eqref{eqs_6b_n2a} becomes negligible for the effective growth rate and the second term of $\Gamma$ in (\ref{gam}) will be dropped.

\par Consequently, from (\ref{eqs_6b_n2a}), we approximate the effective growth rate of perturbations in film thickness by
\begin{equation}\label{eqs_6b_n3_disp_thick}
\frac{d\lambda_{h,r}}{dt}\approx\Lambda \Gamma+\left[\dfrac{h_b^3\phi\left(\alpha h_b\right)}{3\phi(\alpha)\left(1+\alpha h_b\right)}\left(\frac{\alpha^2}{\varsigma\left(1+\alpha h_b\right)^2}-k^2\right)\right]k^2, 
\end{equation}
where
$h_b$ is determined by solving the first equation of the $O(1)$ equation (\ref{eqs_6c_jj}), and $c_b$ in $\Gamma$ is obtained from the second equation of (\ref{eqs_6c_jj}). For $\Lambda=0$ and $h_b=1$, that is, for the case without vapor absorption, the effective growth rate (\ref{eqs_6b_n3_disp_thick}) exactly recovers equation (\ref{eqs_4}). 

\begin{figure}
\includegraphics[scale=0.3]{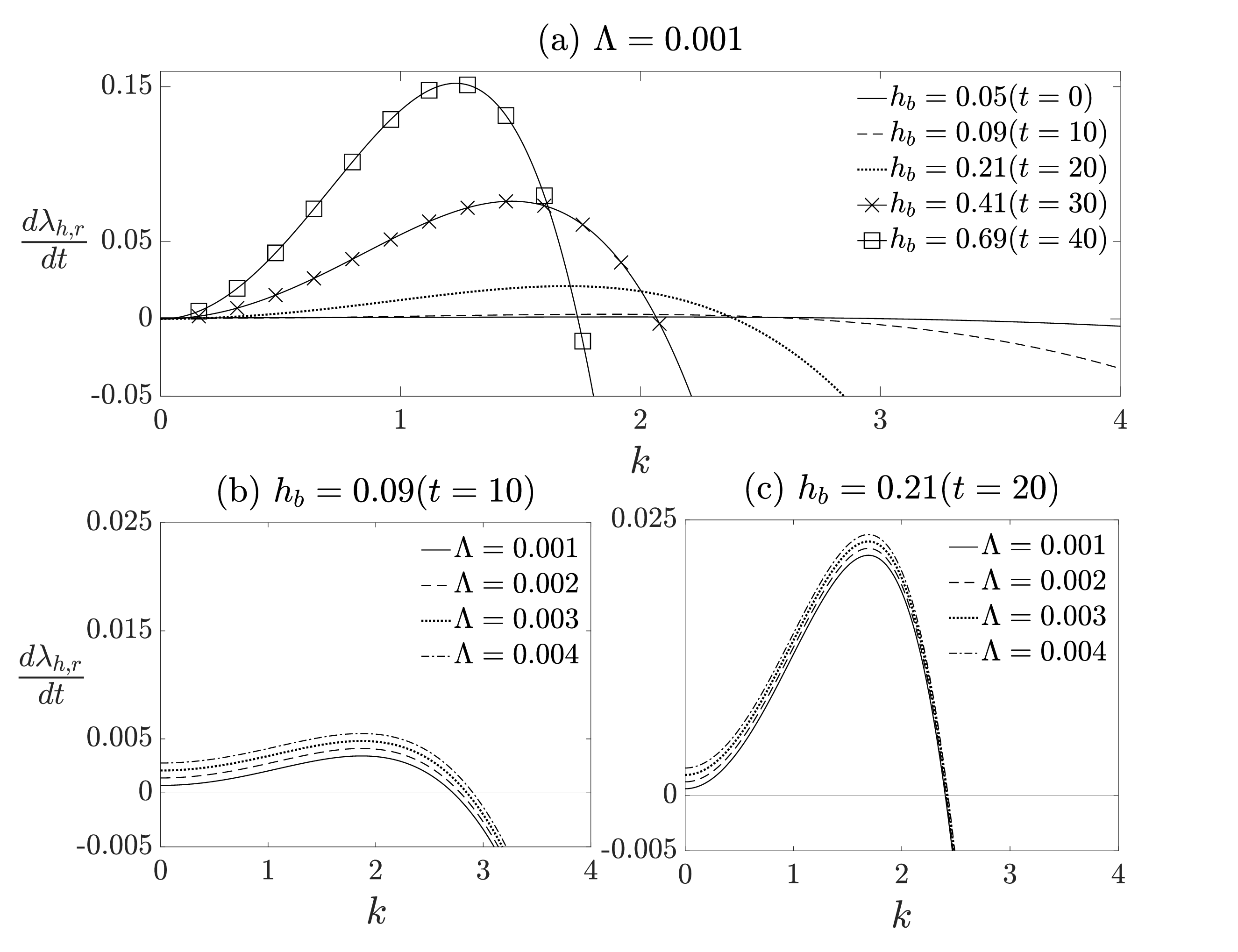} 
\caption{Linear growth rate vs. wavenumber for different values of (a)  $h_b$; 
(b,c)  $\Lambda$. }
\label{fig4}
\end{figure}

\par Figure \ref{fig4} illustrates the effect of weak absorption rate $\Lambda$ and base film thickness $h_b$ on the linear growth rate curves as the wavenumber $k$ varies. In Figure \ref{fig4}a, the linear growth rate's dependence on $k$ is shown for different values of $h_b$ when absorption is present ($\Lambda = 10^{-3}$). We have selected five representative values of $h_b$: 0.05, 0.09, 0.21, 0.41, and 0.69, corresponding to times $t = 0, 10, 20, 30, 40$, respectively, based on solutions to the ODE \eqref{eqs_6c_jj}. The results indicate that increasing the base film thickness intensifies the flow instability but over a narrower range of wavenumbers. This observation is consistent with  \cite{ji2019dynamics}. To examine the effect of $\Lambda$ on the linear growth rate, Figures \ref{fig4}b and \ref{fig4}c display the results at two values of $h_b$: $h_b = 0.09$ and $h_b = 0.21$, corresponding to times $t = 10$ and $t = 20$, respectively, from Figure \ref{fig4}a. Four typical values of $\Lambda$ are considered: $\Lambda = [1-4] \times 10^{-3}$ to explore its impact. The findings show that a higher absorption rate broadens the wavenumber range, enhancing flow instability. However, comparing the figures in the bottom panel, we observe that higher absorption increases flow instability for larger values of $h_b$, but the range of wavenumbers remains almost unchanged.

\par Figure \ref{fig4} illustrates the existence of a critical wavenumber $k=k_c$ for the base liquid thickness $\left(h_b\right)$ and the absorption parameter $(\Lambda)$, where $0<k<k_c$ results in linear instabilities. To determine this $k_c$, we set $d\lambda_{h,r}/dt=0$ in  (\ref{eqs_6b_n3_disp_thick}) and this yields
\begin{equation}\label{eq_sdip_crit}
k_c=\sqrt{\frac{\mathbb X+\sqrt{\mathbb X^2+4\mathbb S\Lambda\Gamma}}{2\mathbb S}}, ~\text{where}\quad
\mathbb X=\frac{\mathbb {S}\alpha^2}{\varsigma\left(1+\alpha h_b\right)^2}, \quad \mathbb S=\dfrac{h_b^3\phi\left(\alpha h_b\right)}{3\phi(\alpha)\left(1+\alpha h_b\right)}.
\end{equation}

\par The expression (\ref{eq_sdip_crit}) demonstrates that the critical wavenumber is affected by the base thickness $h_b(t)$, base concentration $c_b(t)$, and the absorption parameter $\Lambda$.
\begin{figure}
\centering
\includegraphics[scale=0.18]{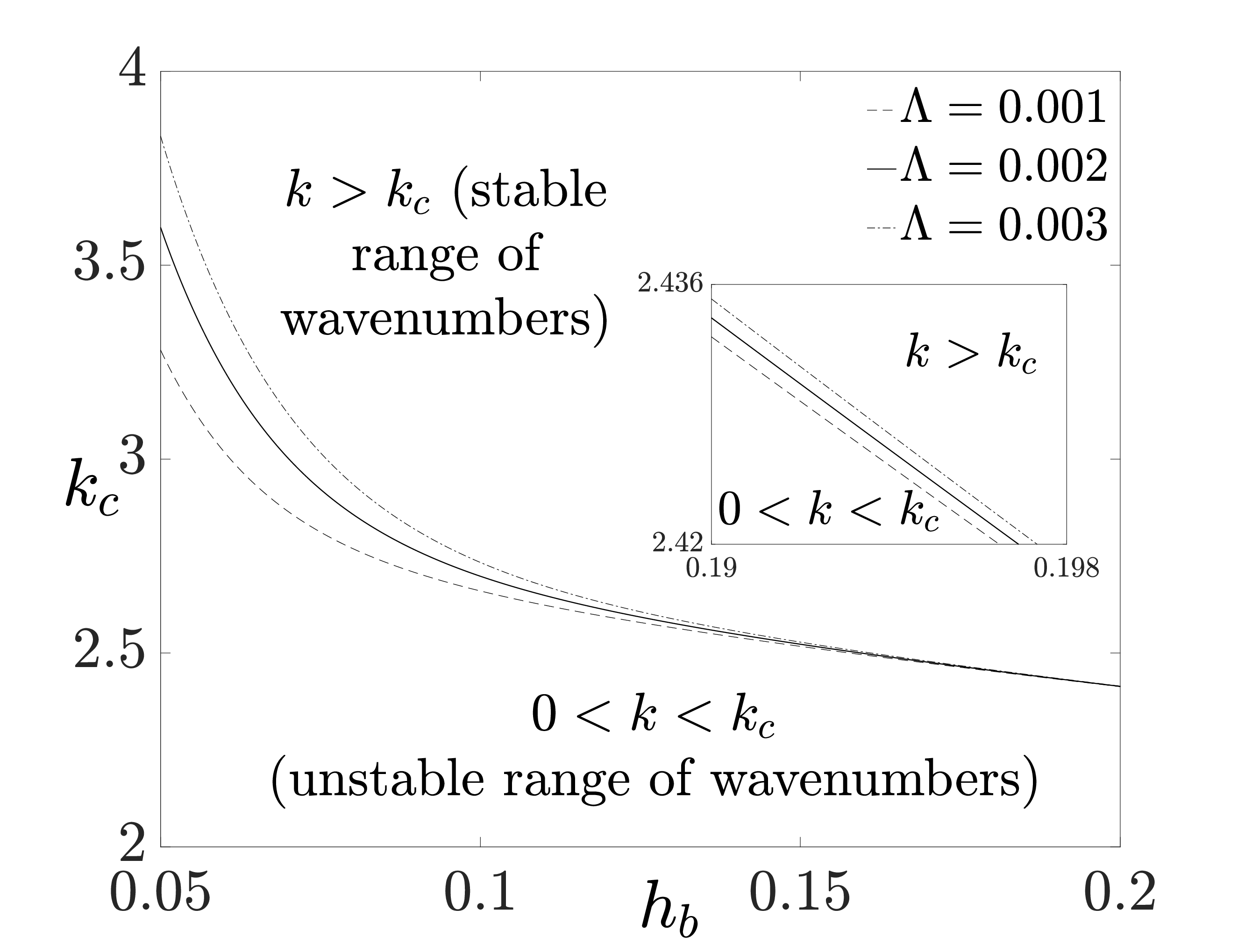} 
\caption{Neutral stability curve as a function of $h_b$ corresponding to different $\Lambda$. }
\label{fig4_n}
\end{figure}
In Figure \ref{fig4_n}, we present the neutral stability curve as a function of $h_b$ for different values of $\Lambda$. The figure demonstrates that, for a fixed film thickness $h_b$, an increase in $\Lambda$ broadens the unstable region. Conversely, for a fixed $\Lambda$, the unstable range of wavenumbers decreases as $h_b$ increases. Additionally, we observe that for thinner films, the differences between the curves corresponding to different values of $\Lambda$ are more pronounced, while for thicker films, the curves nearly overlap. These findings align with the results shown in Figure \ref{fig4}.

\par Moreover, we obtain the most unstable mode $k=k_m$ by setting the derivative of $d\lambda_{h,r}/dt$ with respect to $k$ to zero. This gives $k_m={\alpha}/[{\sqrt{2\varsigma}\left(1+\alpha h_b\right)}],$ which is independent of the absoption parameter $\Lambda$. For the special case $\Lambda = 0$ (without absorption), the expressions for the critical wavenumber $k_c$ and the most unstable mode $k_m$ reduce to those obtained for mass-conserving fibre coating systems \citep{liu2017stability, halpern2017slip,ji2018instability}, except that in those cases, the base state $h_b$ is assumed to be static.

\begin{figure}
\centering
\subfloat[]{
\includegraphics[width=6.5cm]{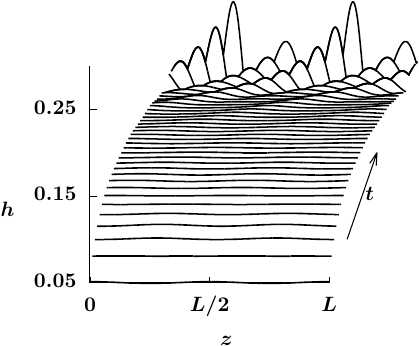}}
\qquad
\subfloat[]{\includegraphics[width=5.5cm]{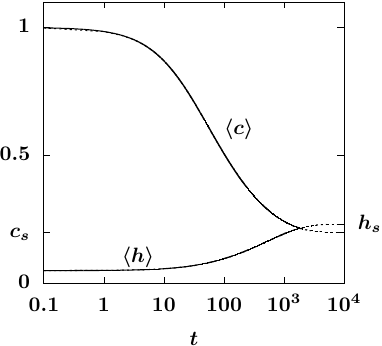}}
\caption{(a) PDE simulation of  \eqref{eq:main} starting from the initial data \eqref{eq:stability_ic} over a periodic domain showing the elevated average film thickness $\left<h\right>$ over time, which evolves into traveling droplets. (b) The average film thickness $\left<h\right>$ and the average concentration $\left<c\right>$ (solid curves) over time, compared against predictions from the ODE \eqref{eqs_6c_jj} (dashed curves). The system parameters are $L = 20$,  $Ma = 20$, $\Lambda=10^{-3}$. 
}
\label{fig:3d_stability}
\end{figure}

\begin{figure}
\centering
\subfloat[]{\includegraphics[height=5cm]{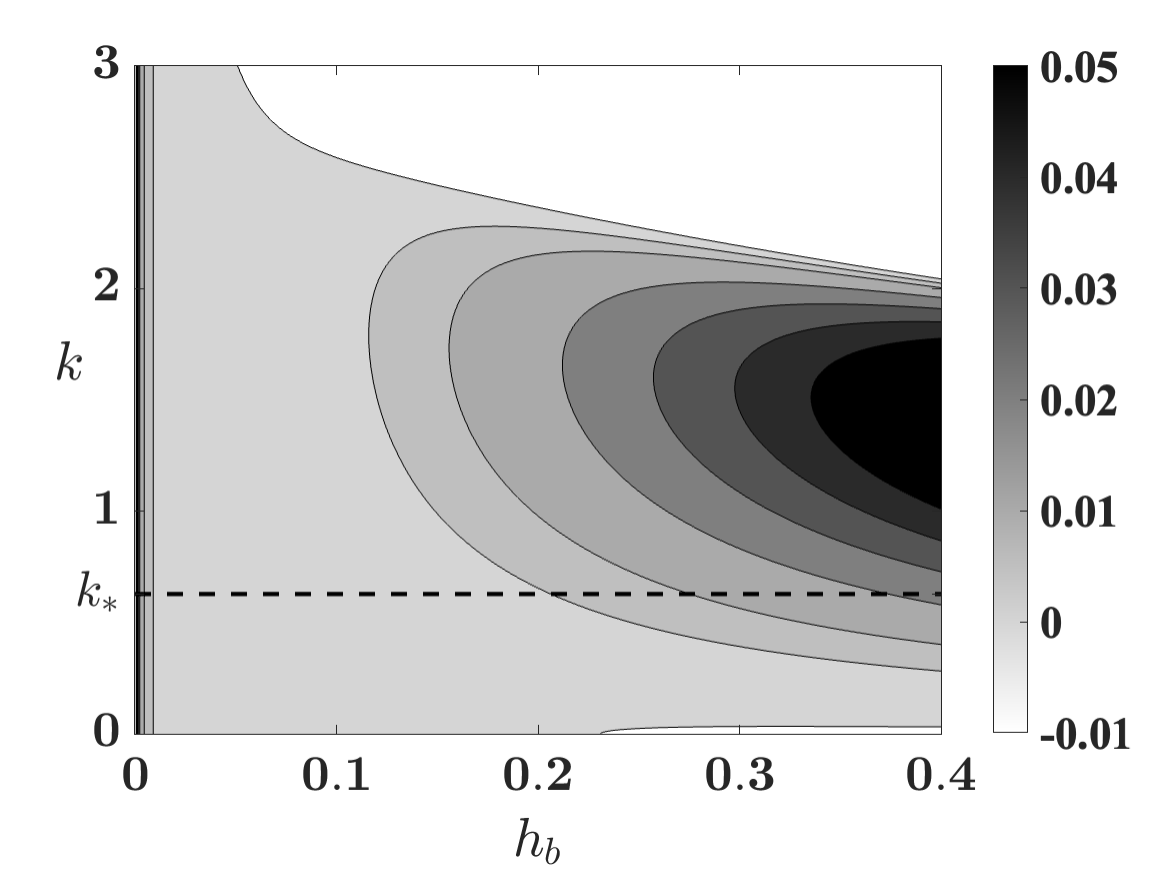}}
\qquad
\subfloat[]{\includegraphics[height=5cm]{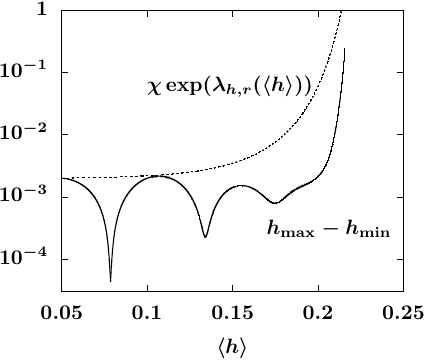}}
\caption{(a) Contour plot of the effective growth rate ${d\lambda_{h,r}}/{dt}$ in \eqref{eqs_6b_n3_disp_thick}, showing that for $k=k_* = 4\pi/L \approx 0.63$, the flat film is linearly unstable for $0 < h_b < h_s$. (b) Plot of $h_{\max}-h_{\min}$ over time for the simulation in Figure \ref{fig:3d_stability}a compared against the prediction $\chi \exp(\lambda_{h,r}(\left<h\right>))$ as the average film thickness $\left<h\right>$ increases.
}
\label{fig:stability_contour}
\end{figure}
To provide an illustration of the instability described by \eqref{eqs_6c} and \eqref{eqs_6b_n3_disp_thick}, we consider an example with weak absorption rate $\Lambda = 10^{-3}$ and the initial condition 
\begin{equation}
\label{eq:stability_ic}
h_0(z) = 0.05+\frac{\chi}{2}\cos\left(\frac{4\pi z}{L}\right),\quad c_0(z) = 1+\frac{\chi}{2}\cos\left(\frac{4\pi z}{L}\right),
\end{equation}
over a periodic domain of length $L = 20$, where $\chi = 0.002$. Here, the initial base thickness $h_b(0) = 0.05$, the initial base oil concentration $c_b(0) = 1$, and the wavenumber $k = k_* = 4\pi/L\approx 0.63$. The simulation in Figure \ref{fig:3d_stability}a shows that in the early stage, the film thickness $\left<h\right>$ increases monotonically over time with small spatial perturbations. Figure \ref{fig:3d_stability}b compares the average film thickness $\left<h\right>$ and the average concentration $\left<c\right>$ against predictions from the leading-order ODE in \eqref{eqs_6c_jj}, showing that the trajectories of the base states match well with the prediction.

\par At the later stage, the ODE predicts that the average film thickness and the average concentration approach the terminal height $h_s$ and saturation concentration $c_s$, respectively. However, the spatial perturbations in the PDE solutions grow significantly and eventually evolve into two moving droplets. This observation is consistent with the prediction from the effective growth rate \eqref{eqs_6b_n3_disp_thick}, as depicted in the contour plot in Figure \ref{fig:stability_contour}a. This plot shows that for $k = k_*$, the liquid film is linearly unstable for $0 \le h_b \le h_s$.
Figure \ref{fig:stability_contour}b presents the magnitude of the spatial perturbation in film thickness, quantified as the difference between the maximum and minimum values of the film height over time, $h_{\max}-h_{\min}$, where $h_{\max}(t) = \max_z h(z,t)$ and $h_{\min}(t) = \min_z h(z,t)$. 
By solving the ODE 
\begin{equation}\label{eqs_6b_n3_disp_thick_12}
\frac{d\lambda_{h,r}}{dh_b}=\frac{\Lambda \Gamma+\left[\dfrac{h_b^3\phi\left(\alpha h_b\right)}{3\phi(\alpha)\left(1+\alpha h_b\right)}\left(\dfrac{\alpha^2}{\varsigma\left(1+\alpha h_b\right)^2}-k^2\right)\right]k^2}{\Lambda\left(c_b-c_s\right)}, 
\end{equation}
we obtain the analytical prediction for the spatial disturbances $\chi\exp(\lambda_{h,r}(\left<h\right>))$ as a function of the average film thickness (dashed curve in Figure \ref{fig:stability_contour}b), where $\chi = 0.002$.
This comparison shows that the stability analysis provides a good prediction for the linear growth rate of the spatial perturbations until the PDE solutions become dominated by the nonlinear dynamics during droplet formation.

\subsection{Absolute and convective instability}\label{sec:4c}
To investigate the effect of the water vapor absorption on the absolute and convective (A/C) instability, we multiply (\ref{eqs_6b_n2a}) by $i=\sqrt{-1}$ and neglect the terms associated with $\left(\widehat c/\widehat h\right)\exp(\lambda_c - \lambda_h)$ as per the discussion after (\ref{eqs_7a_n21}). Further, using the following transformation 
\begin{equation}\label{eqs_6b_n3_disp_thick100}
i\left(\frac{d\lambda_{h}}{dt},\Lambda\right)\rightarrow \frac{\mathbb E^{4/3}}{\left(3\mathbb S \right)^{1/3}}\left(\frac{d\lambda_{h}^\dagger}{dt},\Lambda^\dagger\right), \quad k\rightarrow \left(\frac{\mathbb E}{3\mathbb S}\right)^{1/3}k^\dagger
\end{equation}
and dropping the $\dagger$ sign yields the following:
\begin{align}\label{eqs_6b_n3_disp_thick101}
&\frac{d\lambda_h}{dt}=\Lambda\Gamma+k+\frac{ik^2}{3}\left(\beta-k^2\right), \nonumber
\\
&\quad
\beta=\left(\frac{3h_b^2\phi\left(\alpha h_b\right)+\alpha h_b^3\phi'\left(\alpha h_b\right)}{3h_b^3\phi\left(\alpha h_b\right)}\right)^{-2/3}\frac{\alpha^2}{\varsigma\left(1+\alpha h_b\right)^2}.
\end{align}
When $\Lambda = 0$, equation (\ref{eqs_6b_n3_disp_thick101}) aligns with the one derived by \cite{frenkel1992nonlinear}. \cite{duprat2007absolute} determined that flow instability transitions between absolute and convective (A/C) regimes depending on whether $\beta$ is above or below a critical threshold, $\beta_{c} \equiv \left[(9/4)\times\left(-17+7\sqrt{7}\right)\right]^{1/3}\approx1.507$. This leads to the absolute/convective instability threshold for the present study as
\begin{equation}\label{eqs_6b_n3_disp_thick102}
\frac{3h_b^3\phi\left(\alpha h_b\right)}{3h_b^2\phi\left(\alpha h_b\right)+\alpha h_b^3\phi'\left(\alpha h_b\right)}\left(\frac{\alpha^2}{\varsigma\left(1+\alpha h_b\right)^2}\right)^{3/2}>\left[\frac{9}{4}\left(-17+7\sqrt{7}\right)\right]^{1/2}\approx1.8495.
\end{equation}
\begin{figure}
\centering
\includegraphics[scale=0.18]{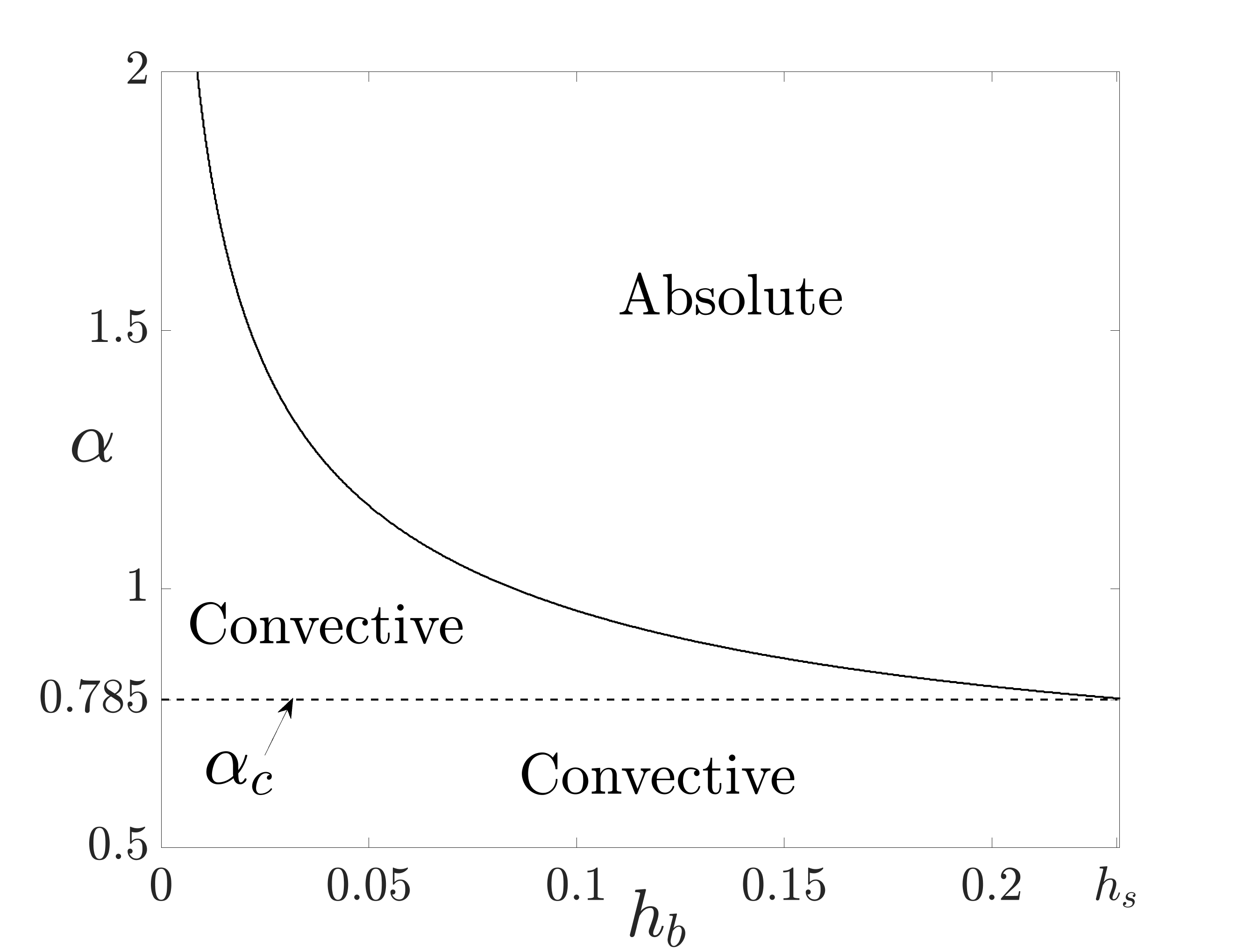} 
\caption{The absolute and convective (A/C) instability regimes from \eqref{eqs_6b_n3_disp_thick102}.
}
\label{fig_abs}
\end{figure}
\par  Equation (\ref{eqs_6b_n3_disp_thick102}) demonstrates that the A/C marginal curve is influenced by the film thickness $h_b$, which, in turn, is affected by the absorption parameter $\Lambda$. As $\Lambda$ increases, $h_b$ also increases (see Figure \ref{fig:base}), thus affecting the A/C marginal curve due to water vapor absorption. In Figure \ref{fig_abs}, we present the A/C instability regimes predicted by (\ref{eqs_6b_n3_disp_thick102}). The figure reveals the presence of a critical value for $\alpha$, denoted by $\alpha_c \approx 0.785$, below which the instability remains convective for any $h_b$. When $\alpha > \alpha_c$, the instability initially appears as convective over a large range of $h_b$, then becomes absolute over a smaller range of $h_b$. Furthermore, for higher values of $\alpha$ beyond $\alpha_c$, the instability remains convective over a smaller range of $h_b$. This study considers $\alpha=0.93$ [see section \ref{sec:3b}], and for this value of $\alpha$, the convective regime transitions to the absolute regime at higher $h_b$ values.

\section{Numerical studies}\label{sec:5}
After gaining initial insights into water absorption from the stability analysis in section \ref{sec:4}, where periodic boundary conditions were applied to study the system's response to perturbations, we now shift to consider more realistic boundary conditions. In this section, we examine the full dynamics of the model (\ref{eq:main}) with Dirichlet boundary conditions at the inlet and Neumann boundary conditions at the outlet. These boundary conditions are more appropriate for capturing the system's behavior in a finite domain, allowing for a detailed exploration of film thickness evolution and concentration under weak and strong absorption effects.

\par For the rest of the study, we focus on the flow pattern in the whole physical domain influenced by spatially-dependent water vapor absorption and Marangoni effects. Following \cite{ji2021thermally,ji2020modelling},
we impose the following Dirichlet boundary conditions at the inlet $z = 0$
\begin{subequations}\label{BCs}
\begin{equation}\label{PDE_BC0}
h(0,t) = h_{\text{IN}},  \quad q(0,t)= q_{\text{IN}}, \quad c(0,t) = 1,
\end{equation}
and the Neumann boundary conditions at the outlet $z = L$,
\begin{equation}\label{PDE_BCL}
h_z(L,t) = 0, \quad h_{zz}(L,t) = 0, \quad c_z(L,t) = 0,
\end{equation}
\end{subequations}
where $h(0,t) = h_{\text{IN}}$ incorporates the nozzle geometry into the system, $q_{\text{IN}}$ is the imposed flow rate, and
$c(0,t)=1$ specifies that the liquid entering the domain is purely silicone liquid sorbent. As vapor absorption occurs, the oil concentration is expected to decay downstream along the fibre. The initial conditions for the film thickness $h$ and the oil concentration $c$ as
\begin{equation}\label{PDE_IC}
h(z,0) = 
\begin{cases}
h_{\text{IN}} + {\left(0.2-h_{\text{IN}}\right)z}/{z_L}  & \text{if}\quad z<z_L,\\
0.2 & \text{if }\quad z\geq z_L.
\end{cases}
\quad \text{and} \quad
c(z,0)\equiv 1.
\end{equation}

\par For all simulations shown in this section, we set $h_{\text{IN}} = 2$, $q_{\text{IN}} = 1/3$, and $z_L = 10$. Based on the available sorption data, the saturated silicone liquid sorbent concentration is set to be $c_s = 0.86$, which corresponds to the saturated water concentration $\eta_s = 1-c_s = 0.14$ (see the discussion in section~\ref{sec:3b}). 
The other parameters are set to
$\varsigma = 0.11$, $\alpha = 0.93$, and $\delta = 0.01$. 
We apply the centered finite difference method and Newton's method to numerically solve the nonlinear PDE model \eqref{eq:main} coupled with the boundary conditions \eqref{BCs} and initial conditions \eqref{PDE_IC}.

\begin{figure}
\centering
\includegraphics[scale=0.32]{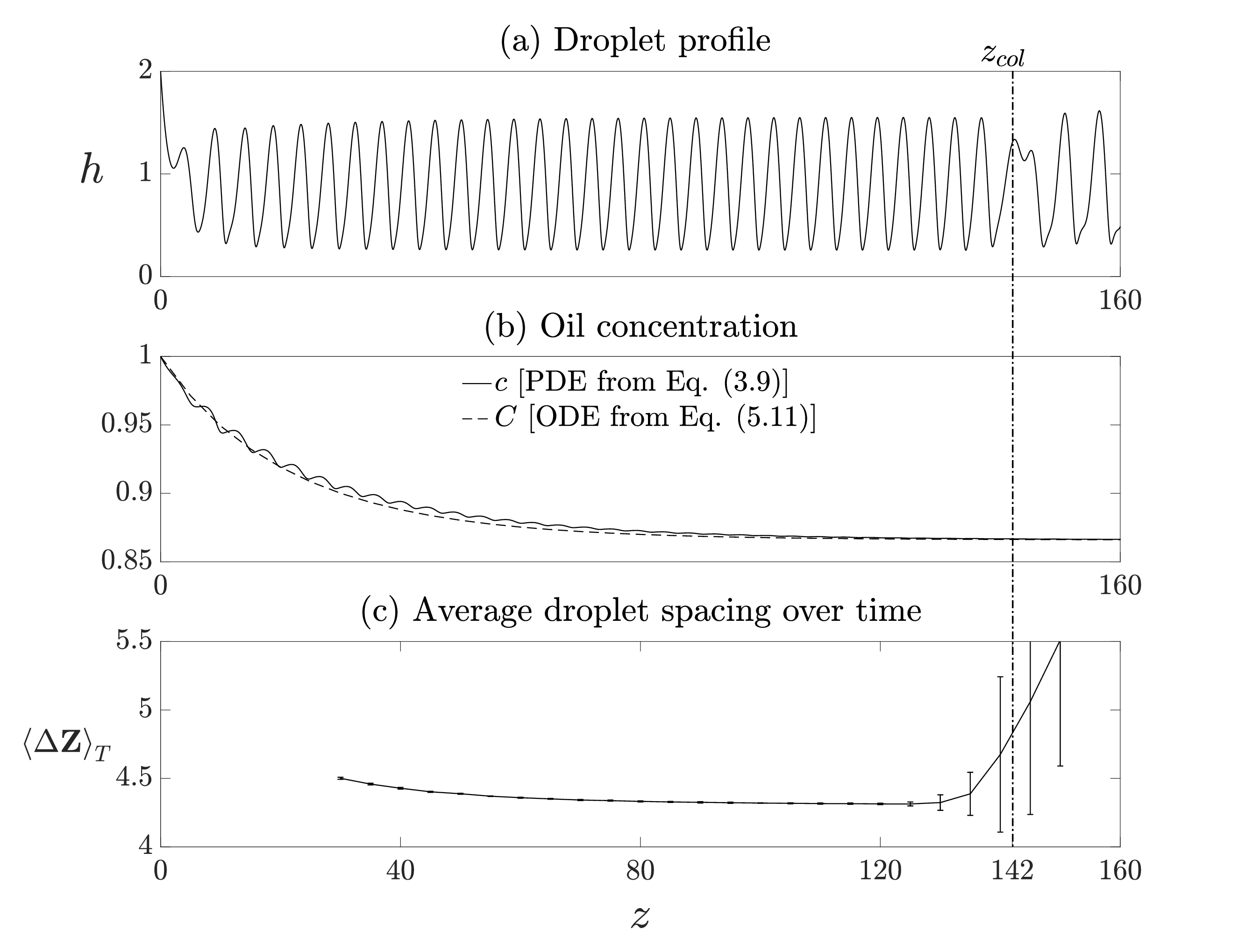}
\caption{Numerical solutions to the PDE model \eqref{eq:main}, subject to the boundary conditions \eqref{BCs} and initial conditions \eqref{PDE_IC}, at time $t = 760$ for (a) the film thickness $h(z,t)$ and (b) the oil concentration $c(z,t)$, compared against the quasi-static profile $C(z)$ in \eqref{swez}, showing that droplet coalescence occurs at $z = z_{col}\approx 142$.
(c) The average droplet spacing $\left< \Delta \mathbf{Z}\right>_T$ as a function of $z$, showing increased variance in droplet spacing downstream for $z > z_{col}$. The system parameters are $\Lambda = 0.0084$ and $Ma = 20$. 
}
\label{Figure 2}
\end{figure}
\par
To numerically capture the droplet dynamics, we observe three characteristics of the droplets: the height of the droplet peaks $\mathbf{H}(z,t)$, the velocity of the droplets $\mathbf{V}(z,t)$, and the spacing between droplets $\Delta \mathbf{Z}(z,t)$, as functions of space $z$ and time $t$. At any time $t$ in the range $T_1\leq t \leq T_2$, there is a train of $N(t)$ droplets in the subdomain $z_1 \le z \le z_2$, where $N(t)$ depends on $t$. The subdomain $z_1 \le z \le z_2$ is selected to ensure that the captured downstream droplet characteristics are minimally influenced by inlet and outlet boundary effects \citep{Ji2020Modeling}.
For the $i^{\text{th}}$ droplet in this subdomain, we denote the corresponding location and the height of the peaks as $\mathbf{Z}_i(t)$ and $\mathbf{H}_i(t)$ for $1\leq i\leq N(t)$. We then define the peak-to-peak distance between the $i^{\text{th}}$ droplet and its right neighbor droplet as
\begin{equation}
   \Delta \mathbf{Z}_i(t) = \mathbf{Z}_{i+1}(t) - \mathbf{Z}_i(t), \quad i = 1, 2, \cdots, N(t)-1.
\end{equation}
The velocity of droplets, $\mathbf{V}_i(t)$, is approximated using finite differences applied to two consecutive snapshots separated by one unit of time.
We define the inter-droplet spacing, height, and velocity functions $\Delta \mathbf{Z}(z,t)$, $\mathbf{H}(z,t)$, and $\mathbf{V}(z,t)$ as
\begin{align}
\label{PSV}
&\Delta \mathbf{Z}(z,t)= \sum_{i=1}^{N(t)-1}\Delta \mathbf{Z}_i(t)\mathds{1}_{\mathbf{Z}_i(t)},\quad
   \mathbf{H}(z,t)= \sum_{i=1}^{N(t)} \mathbf{H}_i(t)\mathds{1}_{\mathbf{Z}_i(t)},\nonumber\\
&\mathbf{V}(z,t)= \sum_{i=1}^{N(t)} \mathbf{V}_i(t)\mathds{1}_{\mathbf{Z}_i(t)},
\end{align}
where $\mathds{1}_{z_1}(z)$ is the indicator function in space defined as
\begin{equation}
   \mathds{1}_{z_1}(z) = 
    \begin{cases}
    1  & \text{if}\quad z = z_1,\\
    0 & \text{if}\quad z\neq  z_1.
\end{cases}
\label{indicator1}
\end{equation}
We then take the time-average of $\Delta \mathbf{Z}(z,t)$, $\mathbf{H}(z,t)$, and $\mathbf{V}(z,t)$ as
\begin{equation}\label{PSVtimeaverage}
   \left<\mathcal X\right>_{T}(z)= \frac{1}{T_2-T_1}\int_{T_1}^{T_2}\mathcal X(z,t)\text{d}t ,
\end{equation}
where $\mathcal X=\Delta \mathbf{Z}, \mathbf{H}, \mathbf{V}$.

\par For all the simulation results presented in this section, we set $z_1 = 20$ and $z_2 = 160$ for the subdomain, and $T_1 = 700$ and $T_2 = 900$ for the time horizon. Numerical observations indicate that this downstream spatial and temporal domain selection yields representative droplet configurations in each flow regime.
The time-averaged quantities defined in \eqref{PSVtimeaverage} for each spatial location $z$ over the subdomain $z_1 \le z \le z_2$  are approximated by numerically tracking the droplet characteristics over a small neighborhood of $z$ and averaging the corresponding quantities over time.

Figure \ref{Figure 2} presents a typical numerical solution of the PDE model \eqref{eq:main} for the film thickness profile and the associated oil concentration where droplet coalescence occurs. This simulation corresponds to the absorption parameter $\Lambda = 0.0084$ and the Marangoni number $Ma = 20$. Figure \ref{Figure 2}a shows the film thickness $h(z,t)$ at $t=760$, where two droplets collide around $z = 142$, forming a larger droplet. Figure \ref{Figure 2}b displays the corresponding oil concentration profile $c(z,t)$ along the domain as a solid line. 
The scatter plot of the average peak-to-peak distance $\left< \Delta \mathbf{Z}\right>_T$ as a function of $z$ is shown in Figure \ref{Figure 2}c. In Figure \ref{Figure 2}b, the concentration decreases to its saturated value $c_s$, which causes $\left< \Delta \mathbf{Z}\right>_T$ to decrease in the range $20<z<130$. This droplet compression behavior ultimately leads to droplet coalescence further downstream. As time progresses, more collisions occur at $z \geq 142$, resulting in an increase in $\left< \Delta \mathbf{Z}\right>_T$ with greater variance, as shown in Figure \ref{Figure 2}c.

\subsection{Quasi-static concentration profiles}\label{sec:5a}
A simplified ODE model can describe the long-time quasi-static profile of the oil concentration $c(z,t)$. Using \eqref{ss1}, we rewrite the PDE \eqref{ss2} as
\begin{equation}
\left(h+\frac{\alpha}{2}h^2\right)c_t + qc_z = \delta\left[c_z\left(h+\frac{\alpha}{2}h^2\right)\right]_z + \Lambda c (1+\alpha h)J.
\label{eq:c_eq}
\end{equation}
We expand the film thickness and the oil  concentration profile as
\begin{equation}
    h(z,t) = 1 + \delta h_1(z,t),\quad c(z,t) = C(z) + \delta c_1(z,t), 
\label{eq:quasi-static_c}
\end{equation}
where the diffusion parameter $\delta \ll 1$. 

\par Substituting  \eqref{eq:quasi-static_c} into \eqref{eq:c_eq} and using \eqref{ss3}--\eqref{swq1}, we obtain the $O(1)$ equation for the leading-order concentration profile $C(z)$ as
\begin{equation}
\left(\frac{1}{3}-Ma\frac{\psi(\alpha)}{\phi(\alpha)}C_z\right) C_z = \Lambda C(z)(1+\alpha) (c_s - C),
\label{eq:leading_order_C}
\end{equation}
where the constant $1/3$  in \eqref{eq:leading_order_C} arises from the mobility function $\mathcal M(h)$ when evaluated at $h=1$.

\par When the water absorption is weak, then $C_z\ll1$, and consequently the $Ma$ term will be dropped from (\ref{eq:leading_order_C}). In this scenario, combining the boundary condition for $c$ in \eqref{PDE_BC0}, equation \eqref{eq:leading_order_C} reduces to a first-order logistic equation for $C(z)$ as
\begin{equation}\label{swez}
\frac{dC}{dz} = 3\Lambda(1+\alpha)C(c_s - C), \quad C(0) = 1,
\end{equation}
which yields the following leading-order profile for the oil concentration:
\begin{equation}\label{swez11}
C(z) = \frac{c_s}{1+\left(c_s-1\right)\exp\left[-3\Lambda(1+\alpha)c_sz\right]}.
\end{equation}
For $z\to\infty$, the quasi-static concentration $C(z)\to c_s$. This suggests that the oil concentration asymptotically approaches the saturation concentration further downstream due to vapor absorption along the domain. 
Figure \ref{Figure 2}b presents a comparison between the quasi-static concentration profile $C(z)$ (dashed curve) in \eqref{swez11} and the PDE solution $c(z,t)$ (solid curve) from a long-time simulation of the PDE \eqref{eq:main}, showing that the quasi-static profile qualitatively captures the decreasing concentration trend over space.

\begin{figure}
\centering
\includegraphics[scale=0.32]{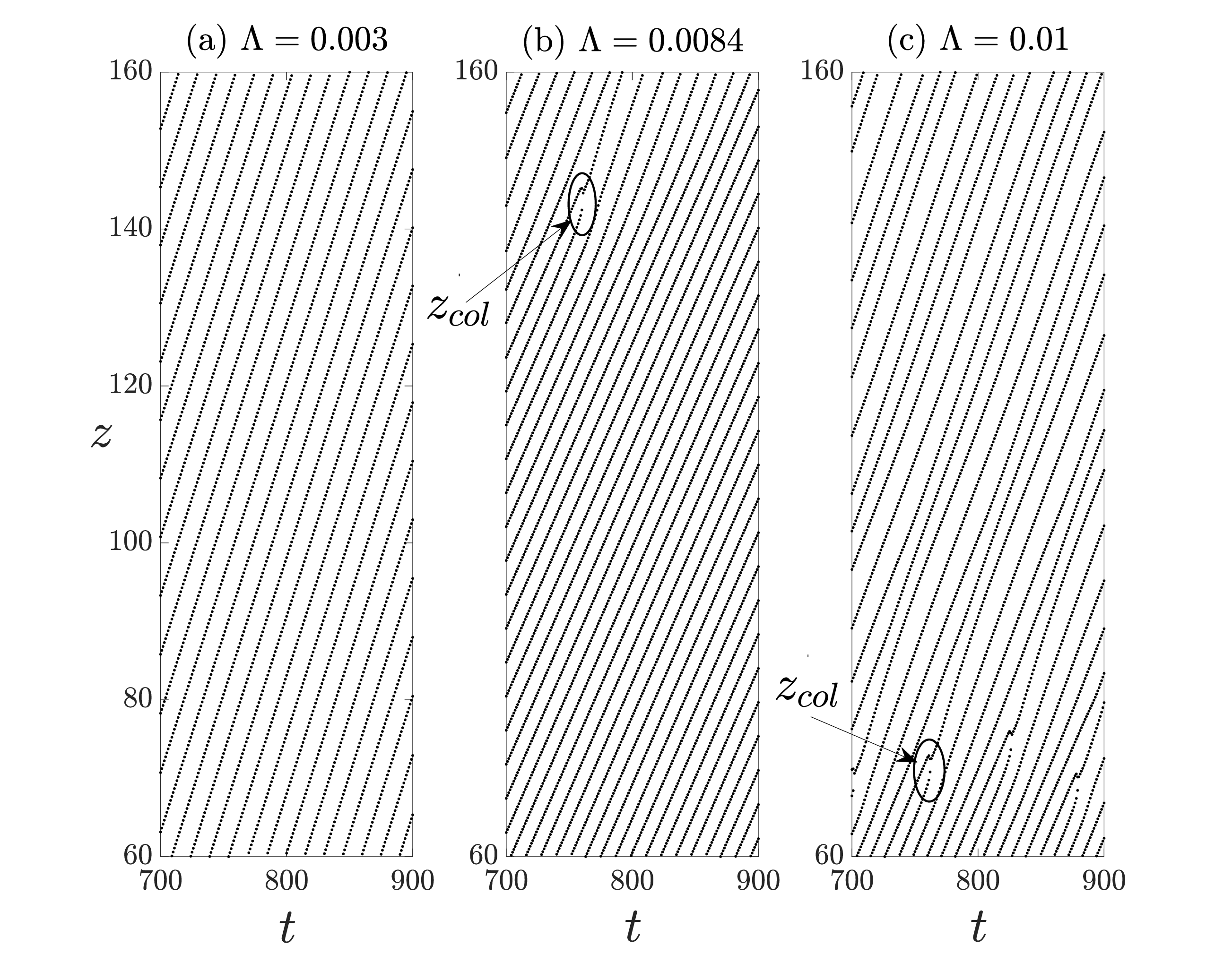}
\caption{Spatiotemporal diagrams for 
(a) $\Lambda = 0.003 < \lambAB$, (b) $\Lambda = 0.0084>\lambAB$, and (c) $\Lambda = 0.01>\lambAB$, showing that droplet coalescence occurs at $z = z_{col}$ for $\Lambda > \lambAB$, and larger $\Lambda$ values can trigger the onset of coalescence closer to the inlet. The other settings are identical to those used in Figure \ref{Figure 2}.}
\label{Figure 1}
\end{figure}

\subsection{Droplet coalescence triggered by absorption}\label{sec:5b}
In this section, we investigate how the droplet coalescence shown in Figure \ref{Figure 2} can be triggered by water absorption. To this end, we present numerical simulations of  \eqref{eq:main}, varying only the $\Lambda$ value while keeping all other parameters identical as in Figure \ref{Figure 2}. 

\par Figure \ref{Figure 1} displays the spatiotemporal diagrams of droplet dynamics for three $\Lambda$ values: $\Lambda = 0.003, 0.0084,  0.01$.
For $\Lambda = 0.003$, as shown in Figure \ref{Figure 1}a, the trajectories of the peaks remain nearly parallel without intersecting. This indicates that the train of droplets moves uniformly to the outlet boundary without any coalescence. This behavior closely resembles the RP regime described by \cite{ji2019dynamics,ji2021thermally}, which we refer to as Regime $\text{I}$.
Figure \ref{Figure 1}b corresponds to the case shown in Figure \ref{Figure 2} for $\Lambda = 0.0084$, where two peaks merge at $z = z_{col} \approx 142$, indicating the occurrence of droplet coalescence. This behavior is similar to the convective regime described by \cite{ruyer2008modelling} and \cite{ji2021thermally}, and we refer to it as Regime $\text{II}$. In Figure \ref{Figure 1}c, for $\Lambda = 0.01$, droplet coalescence occurs at $z = z_{col} \approx 70$, which is closer to the inlet. These numerical results suggest the existence of a threshold value
$\Lambda = \lambAB$ at which a regime transition takes place.
No coalescence occurs when $\Lambda<\lambAB$, as shown in Figure \ref{Figure 1}a. In contrast, coalescence can take place when $\Lambda > \lambAB$, as shown in Figures \ref{Figure 1}b and \ref{Figure 1}c. 
A comparison between Figures \ref{Figure 1}b and \ref{Figure 1}c shows that increased water absorption promotes the onset of droplet coalescence, and stronger absorption leads to coalescence occurring further upstream. Later in this subsection, we will investigate another regime transition from Regime $\text{II}$ to Regime $\text{I}$, at a second threshold value for the absorption parameter $\Lambda = \lambBA$.

\begin{figure}
\centering
\includegraphics[scale=0.26]{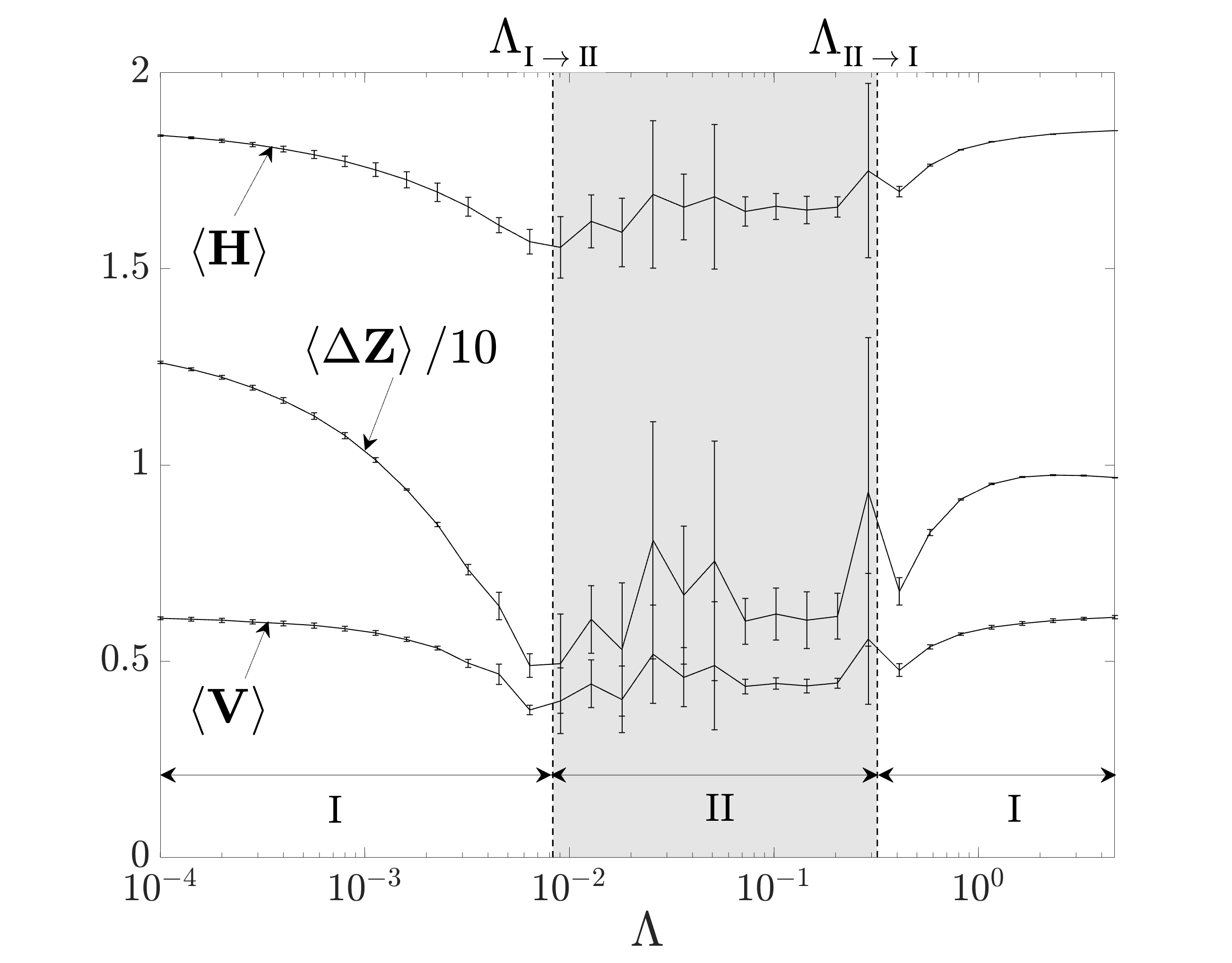}
\caption{Average peak height $\left<\mathbf{H}\right>$, spacing $\left<\Delta \mathbf{Z}\right>$,
and  droplet velocity $\left<\mathbf{V}\right>$ as functions of $\Lambda$ with $Ma = 20$, showing that coalescence occurs for $\lambAB < \Lambda < \lambBA$, where $\lambAB \approx 0.0083$ and $\lambBA\approx 0.32$. All other settings are identical to those in Figure \ref{Figure 2}.}
\label{Figure 3}
\end{figure}

\par To better understand how the regime transition depends on the absorption parameter $\Lambda$, we perform a set of numerical simulations for $10^{-4}\le \Lambda \le 5$ while keeping the other settings identical to those used in Figure \ref{Figure 2}. To capture the global characteristics of the droplets, we also define the spatial-temporal average of the peak-to-peak spacing, droplet height, and velocity by taking the spatial average of the functions $\left< \Delta \mathbf{Z}\right>_{T}(z)$, $\left< \mathbf{H}\right>_{T}(z)$, and $\left< \mathbf{V}\right>_{T}(z)$ defined in  \eqref{PSVtimeaverage}:
\begin{equation}\label{eq:char_average}
    \left<\mathcal X\right> = \frac{1}{z_2-z_1}\int_{z_1}^{z_2} \left< \mathcal X\right>_{T} \text{d}z ,
\end{equation}
where $\mathcal X=\Delta \mathbf{Z}, \mathbf{H}, \mathbf{V}$.

\par In Figure \ref{Figure 3}, we show the dependence of the average droplet peak height $\left< \mathbf{H}\right>$, peak-to-peak spacing $\left< \Delta \mathbf{Z}\right>$, and droplet velocity $\left< \mathbf{V}\right>$ on the absorption parameter, plotted on a semi-logarithmic scale. These spatial-temporal averaged values are numerically obtained by averaging the functions defined in \eqref{PSVtimeaverage} over the subdomain $z_1 \le z \le z_2$. This plot indicates that the droplet dynamics transition from Regime $\text{I}$ to Regime $\text{II}$ at around $\lambAB = 0.0083$, consistent with the transition observed in Figure \ref{Figure 1}, and reenter Regime $\text{I}$ at around $\lambBA = 0.32$. 
The onset of droplet coalescence is only observed for absorption parameter values within the range $\lambAB<\Lambda<\lambBA.$

\par For $\Lambda<\lambAB$, the droplet spacing $\left<\Delta\mathbf{Z}\right>$, height $\left<\mathbf{H}\right>$, and velocity $\left<\mathbf{V}\right>$ decrease with increasing $\Lambda$, indicating a more closely packed droplet configuration with smaller heights and velocities, eventually leading to coalescence when $\Lambda$ reaches $\lambAB$.
The droplet and concentration profiles immediately after this transition $(\Lambda = 0.0084 > \lambAB)$ are presented in Figure \ref{Figure 2}.
In Regime $\text{II}$,  coalescence occurs, and the droplet height $\left< \mathbf{H}\right>$, spacing $\left< \Delta \mathbf{Z}\right>$, and velocity $\left< \mathbf{V}\right>$ become highly sensitive to $\Lambda$, exhibiting large variance. This corresponds to the flow of Regime II dynamics with repeated coalescence events.
For $\Lambda>\lambBA$, the flow reenters Regime $\text{I}$, where a train of non-coalescing droplets flows down the fibre. 
In this case, all three characteristics increase with respect to $\Lambda$, where the increasing average spacing $\left<\Delta\mathbf{Z}\right>$ indicates that the droplets become more separated along the fibre. The average height $\left<\mathbf{H}\right>$ of fully developed droplets increases with the stronger absorption rate, and the corresponding average velocity $\left<\mathbf{V}\right>$ also increases. The comparison between weak absorption ($\Lambda < \lambAB$) and strong absorption ($\Lambda > \lambBA$) cases suggests that absorption can lead to complex trends in droplet configuration due to the non-mass-conserving effects.

\par
\begin{figure}
\centering
\includegraphics[scale=0.32]{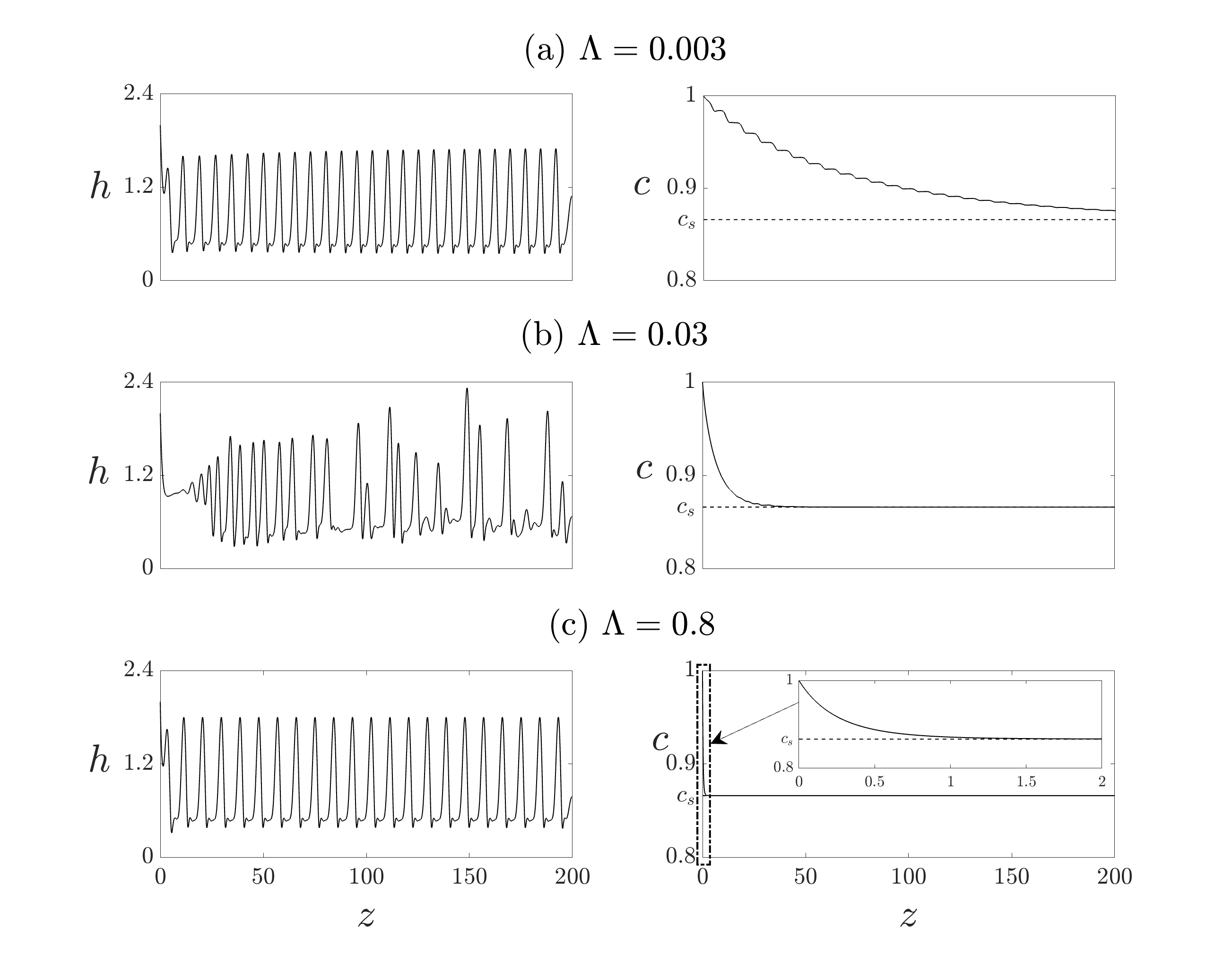}
\caption{Film thickness $h(z,t)$ and oil concentration $c(z,t)$ profiles with representative values of $\Lambda$ in each regime: (a) Regime $\text{I}$ with $\Lambda = 0.003 < \lambAB$; (b) Regime $\text{II}$ with $\lambAB <\Lambda = 0.03 < \lambBA$; (c) Regime  $\text{I}$ with $\Lambda = 0.8 > \lambBA$. The other settings are identical to those in Figure \ref{Figure 2}.}
\label{Figure 4}
\end{figure}

 Figure \ref{Figure 4} displays typical simulation results with representative values of $\Lambda$ in each regime. When $\Lambda = 0.003<\lambAB $, the droplets are in Regime $\text{I}$. They maintain similar heights, move at similar speeds, and have a slow downstream decrease in oil concentration before reaching the outlet boundary. When $\lambAB < \Lambda = 0.03 < \lambBA$,
the droplets are in Regime $\text{II}$, where collisions occur repeatedly, forming larger droplets. The concentration reaches the saturated value $c_s$ at a much faster rate. 
When $\Lambda = 0.8>\lambBA$, coalescence no longer occurs and the droplets move at nearly the same velocity. This indicates a return to Regime $\text{I}$. Additionally, in this case, the oil concentration distribution is more uniform along the domain as it reaches saturation very close to the nozzle inlet.

\begin{figure}
\centering
\includegraphics[scale=0.24]{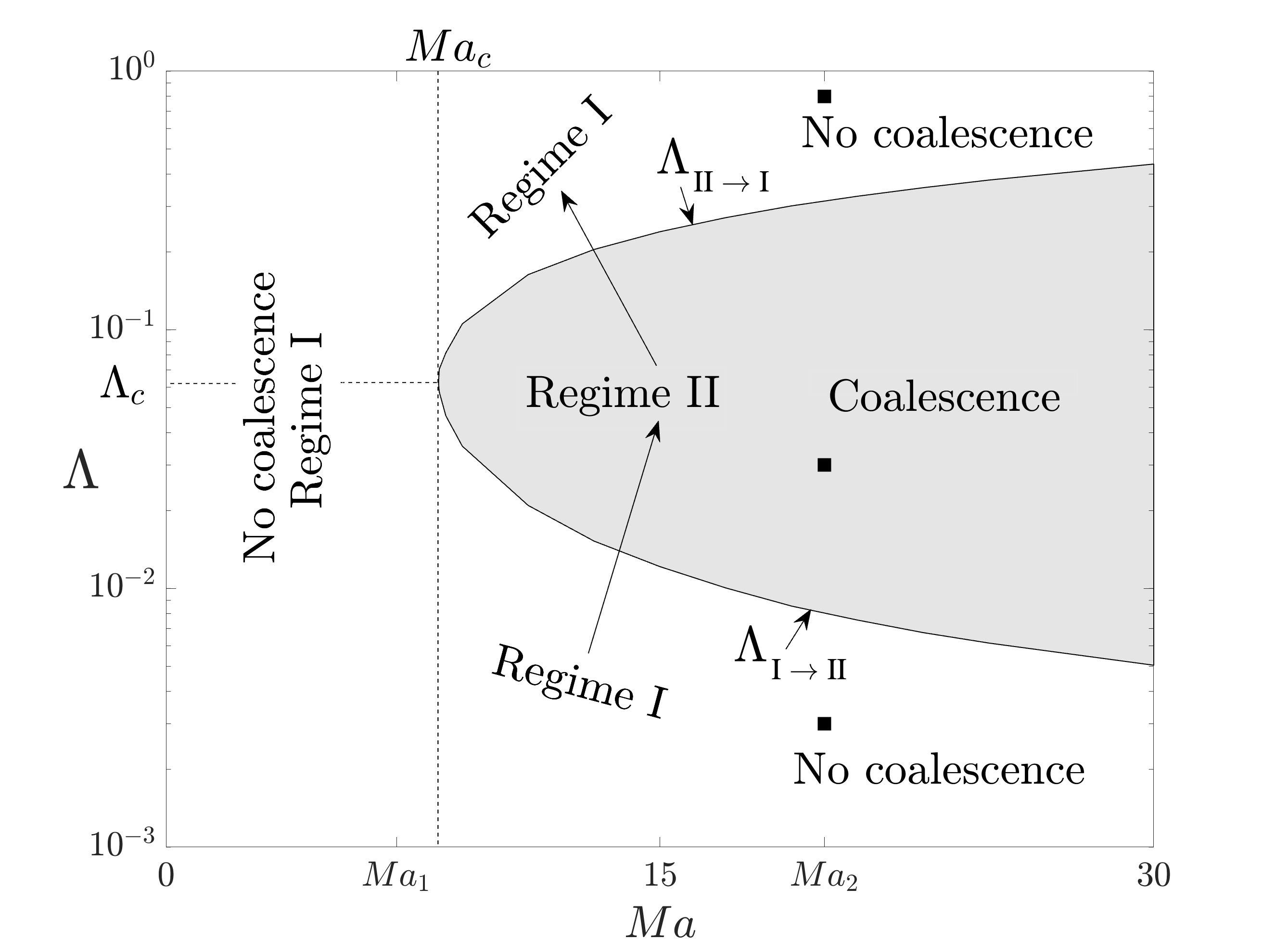}
\caption{Phase diagram of droplet dynamics parameterized by the Marangoni parameter $Ma$ and the absorption parameter $\Lambda$, showing Regime I (no coalescence) and Regime II (coalescence, shaded area).
The critical values are $Ma = Ma_c \approx 8.28$ and $\Lambda = \Lambda_c \approx 0.063$. The three black squares correspond to the numerical results displayed in Figure \ref{Figure 4}. Two typical values, $0<Ma_1 = 7 < Ma_c$ (Case 2) and $Ma_2 = 20 > Ma_c$ (Case 3), correspond to the average mass plots shown in Figure \ref{Figure 10}. }
\label{Figure 9}
\end{figure}

\subsection{Regime transition influenced by absorption and Marangoni effects}\label{sec:5c}
The droplet regimes are also strongly influenced by the Marangoni effects, which are induced by the concentration gradient. 
This section focuses on the combined effects of the Marangoni number $Ma$ and the absorption rate $\Lambda$ on droplet dynamics. 
In Figure \ref{Figure 9}, we present a phase diagram of Regimes $\text{I}$ and $\text{II}$ for $0 \leq \Lambda \leq 1$ and $0 \leq Ma  \leq 30$. This diagram is obtained by numerically simulating the PDE \eqref{eq:main} with varying values of $Ma$ and $\Lambda$. 
For simulations with a specific value of $Ma$, we set the increment in $\Lambda$ to $10^{-3}$ for $\Lambda \leq 0.02$ and $10^{-2}$ for $\Lambda > 0.02$, and numerically observe the droplet dynamics until time $t = T_2 = 900$.
The shaded area in the diagram indicates Regime II, where droplet coalescence occurs, while the unshaded area indicates Regime I, where no coalescence takes place. 
The two regimes are separated by two threshold curves, $\lambAB$ and $\lambBA$, both of which are functions of $Ma$ (see Figure \ref{Figure 9}).
For a given value of $Ma$, these curves provide the two threshold values of $\Lambda$ for the transitions between Regime $\text{I}$ and Regime $\text{II}$, as discussed previously (see Figure \ref{Figure 3} for an example with $Ma = 20$).  These two curves meet at a critical Marangoni number $Ma = {Ma}_c\approx 8.28$, $\lambAB(Ma_c) = \lambBA(Ma_c) = \Lambda_c$, with the corresponding critical absorption parameter $\Lambda = \Lambda_c\approx 0.063$. 

\par Based on the observations, we summarize three cases for droplet dynamic regimes:
\begin{itemize}
\item ~Case 1:  $\Lambda = 0$ (no water vapor absorption): In this case, the concentration $c$ remains constant over time, i.e., $ c(z,t) \equiv C_0$.
The droplets remain in Regime $\text{I}$, with the average spacing
$\left<\Delta \mathbf{Z}\right>$ is relatively constant across the domain, and no coalescence occurs. 
\item ~Case 2: $\Lambda > 0$ and $ 0<Ma<Ma_c$  (weak Marangoni effects): In this case, the concentration profile $c$ decays over space (similar to those in Figure \ref{Figure 4}), but droplet coalescence does not occur and the droplets remain in Regime $\text{I}$.
For a fixed value of $Ma$, the average spacing $\left< \Delta \mathbf{Z}\right>$ decreases with increasing $\Lambda$, leading to a more packed droplet configuration without coalescence.
\item ~Case 3: $\Lambda>0$ and $ Ma>Ma_c$ (strong Marangoni effects): In this case, the droplet regime depends on the absorption rate $\Lambda$. For a specific value of $Ma$, for $\Lambda < \lambAB$ (weak absorption) and $\Lambda > \lambBA$ (strong absorption), the droplets remain in Regime I without coalescence. For intermediate absorption rates $\lambAB < \Lambda < \lambBA$, droplet coalescence occurs, and the dynamics fall into Regime II (shaded area in Figure \ref{Figure 9}). The droplet characteristics in Figure \ref{Figure 3} describe the influence of $\Lambda$ on the regime transition for this case. The threshold value $\lambAB$ ($\lambBA$) is a decreasing (increasing) function of $Ma$, and we have $\lambAB = \lambBA = \Lambda_c$ for the critical Marangoni number $Ma = Ma_c$.
\end{itemize}

\par Depending on the flow regime, the total amount of liquid contained in the droplets over a fixed spatial region varies as a function of $\Lambda$ in different manners. To study this behavior, we numerically calculate the instantaneous mass of liquid over the subdomain $z_1 \le z \le z_2$ and define the averaged liquid mass over time following the definition of $M_l(t)$ in  \eqref{eq:mass},
\begin{equation}
\left<M_l\right>_T = \frac{1}{T_2-T_1}\int_{T_1}^{T_2}\int_{z_1}^{z_2} \left(h+\frac{\alpha}{2}h^2\right) ~\text{d}z~\text{d}t.
\label{eq:mass_average}
\end{equation}
We note that the rate of change of the instantaneous mass is determined by the flux $q$ at $z=z_1$ and $z=z_2$, as well as the non-mass-conserving flux $J$ due to water absorption.
As a result, although the flow rate at the inlet $z = 0$ is fixed, the instantaneous mass is not expected to be conserved over time, and the time-averaged liquid mass $\left<M_l\right>_T$ is expected to highly depend on the flow regime.

\par
\begin{figure}
\centering
\includegraphics[scale=0.283]{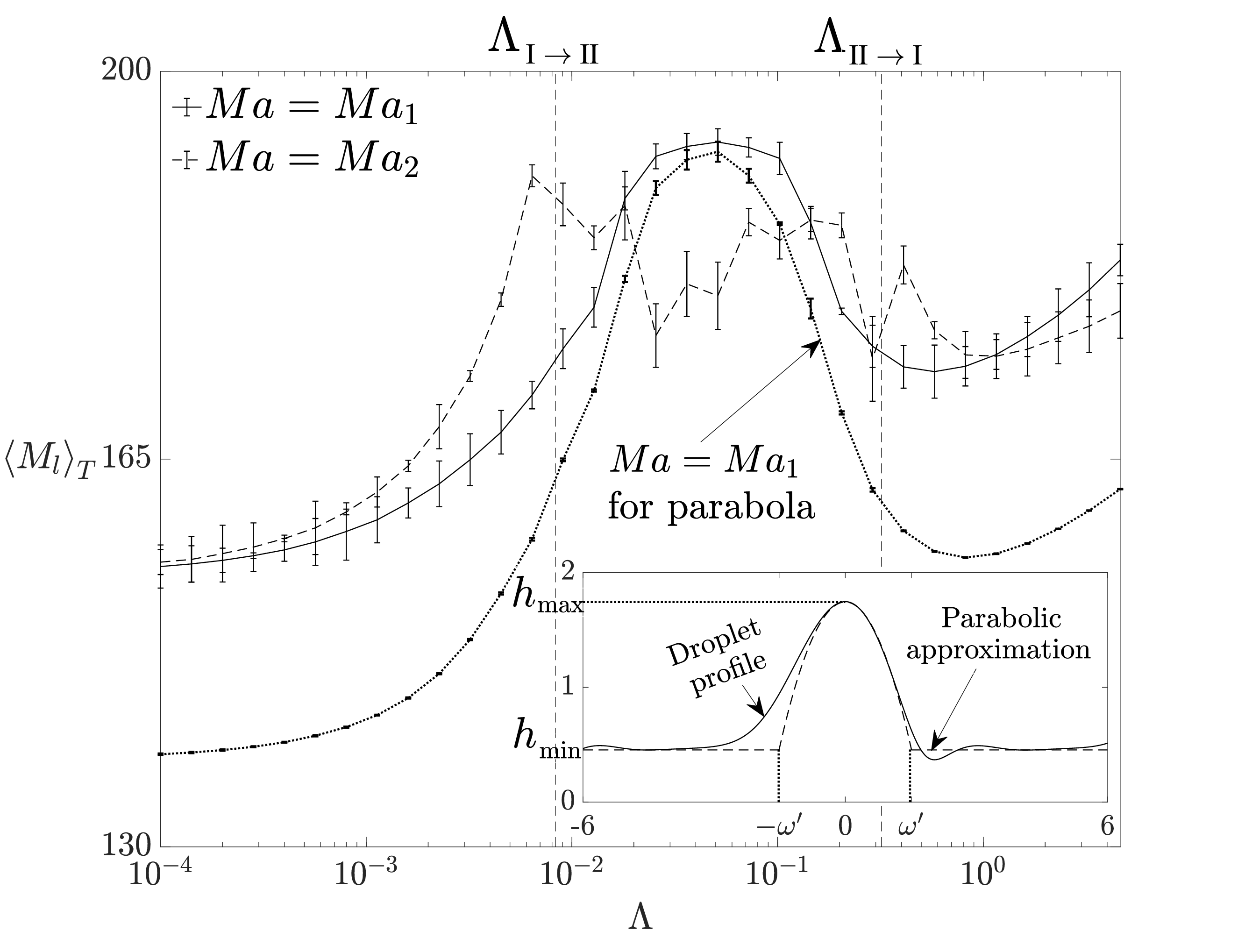}
\caption{The total mass of the droplets in the subdomain $z_1\leq z \leq z_2$ averaged in time with $Ma = Ma_1$ and $Ma = Ma_2$ marked in Figure \ref{Figure 9}. The dotted line is for the total mass of the parabolic approximation when $Ma = Ma_1$. $\Lambda_{\text{I} \rightarrow \text{II}}$ and $\Lambda_{\text{II} \rightarrow \text{I}}$ are the transition threshold for $Ma_2$ only, and they are the same as those in Figure \ref{Figure 3}. The subplot shows the parabola approximation compared with the real droplet profile.}
\label{Figure 10}
\end{figure}

 
In Figure \ref{Figure 10}, we present the averaged total liquid mass $\left<M_l\right>_T$ in the subdomain $z_1 \le z \le z_2$ for two typical values of $Ma$, with $Ma_1 = 7 <{Ma}_c$ (Case 2) and $Ma_2 = 20 > {Ma}_c$ (Case 3), which are marked in Figure \ref{Figure 9}. 
The thresholds
$\Lambda_{\text{I} \rightarrow \text{II}}$ and $\Lambda_{\text{II} \rightarrow \text{I}}$ for regime transitions with $Ma = Ma_2$ are also marked by two vertical dashed lines
in Figure \ref{Figure 10}. We observe that while $Ma = Ma_1$ and $Ma = Ma_2$ lead to significantly different droplet dynamics in the intermediate absorption rate regime, the trends of $\left<M_l\right>_T$ in the weak and strong absorption rate limits for $Ma = Ma_1$ and those of $Ma = Ma_2$ are similar. Specifically, in the weak absorption limit, the mass $\left<M_l\right>_T$ decreases as $\Lambda \to 0^+$; In the strong absorption limit as $\Lambda \to \infty$, the mass $\left<M_l\right>_T$ increases as the absorption rate increases.

\par To understand the trend in the total mass as the absorption rate varies, we approximate the total mass as a function of the average droplet height $\left<\mathbf{H}\right>$ and average droplet spacing $\left<\Delta \mathbf{Z}\right>$. Although the droplet profiles are asymmetric due to the nature of gravity-driven flows, we neglect gravity and surface tension gradient in this approximation and assume that the droplets are quasi-static and symmetric.

\par We consider a quasi-static symmetric droplet placed at the origin $z = 0$.
By setting $\Lambda = Ma = 0$ in  \eqref{eq:main} and neglecting the gravity contribution, we obtain the ODE for a quasi-static symmetric droplet profile $\bar{H}(z)$,
\begin{equation}
\frac{\alpha}{\varsigma(1+\alpha \bar{H})}-\bar{H}_{zz} = p,
\label{eq:quasiStatic}
\end{equation}
where the pressure $p$ is a constant [see pressure solution in (\ref{eq25})].
The ODE (\ref{eq:quasiStatic}) admits a homoclinic droplet profile with its peak $\bar{H}=\hmax$ and the associated precursor layer thickness $\bar{H} = h_{\min}$. Integrating \eqref{eq:quasiStatic} once with respect to $z$ and using the conditions $\bar{H}_z = 0$ when $\bar{H} = \hmax$ and $\bar{H} = h_{\min}$ gives
\begin{equation}
p=\frac{1}{\varsigma\left(\hmax-h_{\min}\right)}\ln\left(\frac{1+\alpha\hmax}{1+\alpha h_{\min}}\right).
\end{equation}

\par Following  \cite{Witelski2003PRE} and \cite{ji2024coarsening}, we approximate the core of the droplet by a parabola
\begin{equation}
\bar{H}(z) \sim A(\omega^2-z^2) \sim -Az^2 + \hmax,
\label{parabola0}
\end{equation}
where the constant $A > 0$, and $\omega$ is the half-width of the droplet Matching the peak of the parabola with the peak of the droplet yields $\bar{H}_{\text{max}}(z) = \bar{H}(0) \sim A\omega^2 = \hmax$.  A sample droplet with its parabolic approximation is shown in the subplot of Figure \ref{Figure 10}. In the core region of the droplet, we approximate the curvature $\bar{H}_{zz}$ as $\bar{H}_{zz} \approx \frac{\alpha}{\varsigma(1+\alpha \hmax)}-p$ in \eqref{eq:quasiStatic}, which combined with the parabolic approximation \eqref{parabola0},
leads to
$A = \left(p-\frac{\alpha}{\varsigma(1+\alpha \hmax)}\right)/2.$
Using \eqref{parabola0}, we also define the effective half-width of the droplet as the half-width of the droplet $\omega'$ at the precursor layer thickness level $h = h_{\min}$, $\omega' = \sqrt{\left(\hmax-h_{\min}\right)/A}$. The subplot in Figure \ref{Figure 10} presents the comparison between a droplet profile obtained from the PDE simulation and its parabolic approximation, showing that the parabolic approximation captures the core of the droplet $-\omega' \leq 0 \leq \omega'$. Here, we pick the minimum of the parabolic approximation $h_{\text{min}}$ at the edges $z = \pm \omega'$ so that it matches the thickness of the precursor layer.

\par Using these estimates, we approximate the mass of a single droplet using a function parametrized by the peak height $\hmax$ as
\begin{subequations}\label{massofDroplet}
\begin{equation}
    M_{\text{drop}}(\hmax;h_{\text{min}}) = \int_{-\omega'}^{\omega'}\left(\bar{H}(z)+\frac{\alpha}{2}\bar{H}^2(z)\right)\text{d}z = \mathbb{F}(\hmax;h_{\min})\omega',
\end{equation}
where 
\begin{equation}
    \mathbb{F}(\hmax;h_{\min}) =\left(8\alpha\hmax^2+4\alpha\hmax h_{\min}+3\alpha h_{\min}^2 + 20\hmax+10h_{\min}\right)/15.
\end{equation}
\end{subequations}
\par To approximate the overall liquid mass on the subdomain $z_1 \leq z \leq z_2$ where a train of nearly equally-spaced traveling droplets coexist, we consider the mass contribution from both the droplets and the precursor layer and write
\begin{equation}
    \left<M_l\right>_T \approx \left<M_{\text{drop}}\right>N+M_{\text{pre}},
\label{parabolaMass}
\end{equation}
where the average total mass over time, $\left<M_l\right>_T$, is defined in \eqref{eq:mass_average}. The average mass of individual droplets is approximated by $\left<M_{\text{drop}}\right> \approx M_{\text{drop}}\left(\left<\mathbf{H}\right>\right)$, where $\left<\mathbf{H}\right>$ is defined in \eqref{eq:char_average}. 
The number of droplets on the subdomain is approximated by $N = \left(z_2-z_1\right)/\left< \Delta \mathbf{Z}\right>_T$, and the mass of the precursor layer, $M_{\text{pre}}$, is given by
\begin{equation}
    M_{\text{pre}} = \int_{z_2-z_1 - 2N\omega '} \left(h_{\min}+\frac{\alpha}{2}h_{\min}^2\right)\text{d}z = \left(h_{\min}+\frac{\alpha}{2}h_{\min}^2\right)\left(z_2-z_1-2N\omega'\right).
\label{massofPrecursor}
\end{equation}
In \eqref{massofPrecursor}, we have assumed that the precursor layer thickness matches the minimum of the approximated droplet profile $h_{\min}$ and is spatially uniform.

\par Figure \ref{Figure 10} shows a comparison of the numerically obtained
total mass $\left<M_l\right>_T$ (solid curve) for the Marangoni numbers $Ma = Ma_1 = 7$ and its approximation (dotted curve) based on \eqref{parabolaMass}. This comparison shows that the parabolic approximation qualitatively captures the trend in the total mass for both weak and strong absorption regimes. The mass increases monotonically for the weak absorption regime until reaching a peak, after which the mass decreases dramatically with increasing $\Lambda$ before starting to increase again in the strong absorption limit. This is consistent with the trends in total mass from PDE simulations.
In this approximation, we set the precursor layer thickness $h_{\min} = 0.455$ based on numerical observations. This approximation neglects the spatial variation in the liquid film interface between droplets and simplifies the gravity-driven droplet profiles to parabolic shapes. This figure concludes that the change in total mass can be estimated by equation \eqref{parabolaMass} in Regime $\text{I}$ and the trend can be predicted by the average droplet height $ \mathbf{\left<H\right>}$ and spacing $\left<\Delta \mathbf{Z}\right>$ curves in Figure \ref{Figure 3}.

\section{Conclusions}\label{sec:6}
In this study, we have developed a lubrication-type model for the dynamics of thin liquid films flowing down a cylindrical fibre while absorbing water vapor. 
The model consists of a coupled PDE system for the liquid film thickness and the water concentration, featuring the interplay between the substrate geometry, surface tension, Marangoni effects, and vapor absorption. Unlike most existing models 
for liquid flowing down a cylindrical fibre, where the total mass of liquid is conserved, our model accounts for non-mass-conserving situations that give rise to more complex and interesting droplet dynamics. 
Stability analysis based on the frozen-time approach shows that the stability of liquid films of nearly flat thickness and concentration profiles highly depends on the non-mass-conserving effects.  Numerical simulations demonstrate that with Dirichlet inlet boundary conditions, the concentration gradient caused by strong vapor absorption triggers droplet coalescence. We numerically identified the influence of the Marangoni number and absorption rate on the regime transition between Regime I (non-coalescence) and Regime II (coalescence) and characterized the corresponding droplet configurations. With weak absorption, the gradually changing concentration profile leads to an array of moving droplets with non-uniform inter-droplet spacings. For stronger absorption, the water concentration exhibits a more rapid decay near the inlet, and when coupled with a sufficiently large Marangoni effect, droplet coalescence can take place. The onset location of the coalescence depends on the absorption rate, and with stronger absorption, the coalescence tends to occur closer to the nozzle.

\par Our model is relatively simple compared to other models \citep{ji2018instability,burelbach1988nonlinear} for volatile fluid. Neglecting inertial effects, for example, reduces the order of our model but also introduces limitations.
To account for the flow dynamics with low to moderate inertial effects, one may consider deriving a weighted-residual model for a more comprehensive understanding of water-absorbing liquid dynamics. The effects of concentration-dependent viscosity and nozzle geometry are also potential topics of further investigation. 

\section*{Acknowledgements}
 H.J. acknowledges support from the National Science Foundation (NSF) under Grant No. DMS-2309774. The work conducted by Y.S.J. is supported in part by a grant from the Council on Research of the Los Angeles Division of
the University of California Academic Senate. The authors also thank Dr. Dongchan (Shaun) Ahn and Dr. Aaron Greiner at the Dow Chemical Company for providing data on the physical properties of the silicone liquid sorbent.

 \section*{Declaration of Interests}
The authors report no conflict of interest.

\section*{Declaration of generative AI and AI-assisted technologies in the writing process}
During the preparation of this work, the authors used ChatGPT to improve the language and readability of the article. After using this tool, the authors reviewed and edited the content as needed and take full responsibility for the content of the publication.

\bibliographystyle{jfm}
\bibliography{jfm}

\end{document}